
\input phyzzx

\date{May, 1994}
\rightline{UTHEP-275}
\titlepage
\vskip 1cm
\title{On the Genus Expansion in the Topological String Theory}
\author {Tohru Eguchi}
\address{Department of Physics, Faculty of Science, University of
Tokyo, Tokyo 113, Japan}
\author{Yasuhiko Yamada}
\address{Department of Mathematics, Kyushu University, Hakozaki,
Fukuoka 812, Japan}
\andauthor{Sung-Kil Yang}
\address{Institute of Physics, University of Tsukuba,
Ibaraki 305, Japan}
\abstract{
A systematic formulation of the higher genus expansion in topological
string theory is considered. We also develop a simple way of evaluating genus
zero correlation functions. At higher genera we derive some interesting
formulas for the free energy in the $A_1$ and $A_2$ models.
We present some evidence that topological minimal models associated with Lie
algebras other than the A-D-E type do not have a consistent higher genus
expansion beyond genus one. We also present some new results on
the $CP^1$ model at higher genera.}
\endpage
\overfullrule=0pt

\def\cmp#1{Commun. Math. Phys. {\bf #1}}
\def\pl#1{Phys. Lett. {\bf B#1}}

\def\np#1{Nucl. Phys. {\bf B#1}}
\def\ijmp#1{Int. J. Mod. Phys. {\bf A#1}}
\def\mpl#1{ Mod. Phys. Lett. {\bf A#1}}
\def\Prim#1{{\cal O}_#1}
\def\Descd#1#2{\sigma_#1({\cal O}_#2)}
\def\bra{\langle}
\def\ket{\rangle}

\REF\W{E. Witten, \np{340} (1990) 281}
\REF\DW{R. Dijkgraaf and E. Witten, \np{342} (1990) 486}
\REF\VV{E. Verlinde and H. Verlinde, \np{348} (1991) 457}
\REF\Witt{E. Witten, \cmp{117} (1988) 353; {\it ibid.} {\bf 118} (1988) 411}
\REF\EY{T. Eguchi and S.-K. Yang, \mpl{5} (1990) 1693}
\REF\Vafa{C. Vafa, \mpl{6} (1991) 337}
\REF\DVV{R. Dijkgraaf, E. Verlinde and H. Verlinde, \np{352} (1991) 59}
\REF\K{I. Krichever, \cmp{143} (1992) 415}
\REF\TT{K. Takasaki and T. Takebe, \ijmp{7} Suppl. 1B (1992) 889}
\REF\D{B. Dubrovin, "Integrable Systems and Classification of
2-Dimensional  Topological Field Theories", SISSA preprint
(September, 1992)}
\REF\EKYY{T. Eguchi, H. Kanno, Y. Yamada and S.-K. Yang,
\pl{305} (1993) 235}
\REF\L{A. Losev, "Descendants Constructed from Matter Field in Topological
Landau-Ginzburg Theories Coupled to Topological Gravity", ITEP preprint
(November, 1992)}
\REF\LP{A. Losev and I. Polyubin, ``On Connection between Topological
Landau-Ginzburg Gravity and Integrable Systems'', ITEP preprint
(May, 1993)}
\REF\Wi{E. Witten, Surveys in Diff. Geom. {\bf 1} (1991) 243}
\REF\KS{K. Saito, Publ. RIMS, Kyoto University {\bf 19} (1983) 1231}
\REF\IZ{C. Itzykson and J.-B. Zuber, \ijmp{7} (1992) 5661}
\REF\Y{T. Yoneya, \ijmp{7} (1992) 4015}
\REF\Gins{P. Ginsparg, M. Goulian, M.R. Plesser and J. Zinn-Justin, \np{342}
(1990) 539}
\REF\JY{A. Jevicki and T. Yoneya, \mpl{5} (1990) 1615}
\REF\FKN{M. Fukuma, H. Kawai and R. Nakayama, \ijmp{6} (1991) 1385}
\REF\DVVer{R. Dijkgraaf, E. Verlinde and H. Verlinde, \np{348} (1991) 435}
\REF\Amb{J. Ambj\o rn, L. Chekhov, C.F. Kristjansen and Yu. Makeenko,
\np{404} (1993) 127}
\REF\KNNT{A. Kato, T. Nakatsu, M. Noumi and T. Takebe, ``Topological
String, Matrix Integral and Singularity Theory'', RIMS preprint,
RIMS-934 (July, 1993)}
\REF\Dub{B. Dubrovin, ``Differential Geometry of the Space of Orbits
of a Coxeter Group'', SISSA preprint (February, 1993)}
\REF\NY{S. Nishigaki and T. Yoneya, \np{348} (1991) 787}
\REF\DKO{P. Di Vecchia, M. Kato and N. Ohta, \np{357} (1991) 495}
\REF\AMP{A. Anderson, R.C. Myers and V. Periwal, \np{360} (1991) 463}
\REF\GD{I.M. Gelfand and L.A. Dikii, Russian Math. Surveys
{\bf 30} (1975) 77}

\chapter{Introduction}

The topological field theory approach to two-dimensional
gravity [\W,\DW,\VV] has uncovered
a fascinating interplay among the twisted
$N=2$ supersymmetry [\Witt,\EY], the theory of KP hierarchy and the
topological Landau-Ginzburg description [\Vafa,\DVV].
It is now well-understood that the integrable structure of the topological
strings at genus zero is governed by the dispersionless KP
hierarchy [\K,\TT,\D]. Physical spectra and recursion relations among
correlation functions of topological strings are described in a simple
manner using the Landau-Ginzburg formulation [\DVV,\EKYY,\L].

There remain at least two issues which should be
better understood in the topological field theory approach to
two-dimensional gravity.
One is how to take into account the full
phase space coupling constants in the Landau-Ginzburg description of
topological strings (for an attempt, see [\LP]).
The other is the explicit and hopefully systematic
evaluation of the higher genus expansion. Studying the latter issue is
our purpose in the present paper. In particular explicit calculations
may be useful since there has been no satisfactory formulation so far
to take into account the higher genera in topological Landau-Ginzburg
approach.

We are motivated by the remarkable expression for the genus one free
energy $F_1$. It has been known that
$$
F_1= {1 \over 24} \log \det u_{\alpha \beta},
\eqn\genusone
$$
where $u_{\alpha \beta}=\partial^3 F_0 /\partial t_0\partial t_\alpha
\partial t_\beta$ with $t_\alpha$ and $F_0$ being the small phase space
couplings and the genus zero free energy, respectively [\DW,\Wi].
It is remarkable
that \genusone\ is valid for topological
gravity coupled to any topological minimal matter.
Moreover \genusone\ gives rise to the genus one correlation
functions expressed entirely in terms of genus zero quantities. In what follows
we shall study if this kind of structure emerges also in the higher genus
free energies.

The paper is organized as follows. In sect.2 we briefly review the genus
zero theory. Some useful formulas for the genus zero free energy
are obtained. Sect.3 is devoted to the higher genus calculations.
The explicit higher genus results for the $A_1$ and $A_2$ models are
presented. In sect.4 we shall examine topological gravity coupled to
topological matter other than the $A$-type models. We present some
evidence against the consistency of models associated with Lie algebras
other than the A-D-E type at higher genus.
We also present some new results on the
higher genus expansion of the $CP^1$ model. Finally in sect.5
we discuss the vector model in the light of our strategy. We relegate to
appendices A, B and C the explicit forms of constitutive relations
for the $A_1$, $A_2$ and $D_4$ models, respectively. In appendix D the
free energy of pure gravity theory up to genus five is obtained.

\chapter{Topological Strings at Genus Zero}

In this section topological gravity coupled to topological matter at genus
zero is considered. We start with reviewing the basic ingredients of the
genus zero theory on the basis of [\DW].
We then discuss how to calculate explicitly
genus zero correlation functions in the full phase space.

Consider a topological matter theory whose BRST invariant observables are
denoted as ${\cal O}_\alpha$ with $\alpha \in I$; a certain set of
integers. In particular ${\cal O}_0$ stands for the
identity operator. In the minimal topological matter, for instance, the
relation between the set $I$ and the exponents of the ADE Lie
algebra is well known. Two-point functions are given by
$$
\langle {\cal O}_\alpha {\cal O}_\beta \rangle =\eta_{\alpha \beta},
\eqn\metric
$$
where $\eta_{\alpha \beta}$ is a topological metric which is non-degenerate
$ \eta_{\alpha \beta}\eta^{\beta \gamma}=\delta_\alpha^\gamma$.
The primary fields ${\cal O}_\alpha$ generate the commutative associative
algebra
$$
{\cal O}_\alpha {\cal O}_\beta=
{C_{\alpha \beta}}^{\gamma} {\cal O}_\gamma.
\eqno\eq
$$
It is clear that
$$
\langle {\cal O}_\alpha {\cal O}_\beta {\cal O}_\gamma \rangle
=C_{\alpha \beta \gamma},
\eqno\eq
$$
where $C_{\alpha \beta \gamma}= {C_{\alpha \beta}}^{\rho }
\eta_{\rho \gamma}$ and $C_{\alpha \beta 0}=\eta_{\alpha \beta}$.

Let us turn on topological gravity. Coupling to topological gravity makes
the physical spectrum of the theory quite rich. In addition to the
original observables ${\cal O}_\alpha$ which we now call gravitational
primary fields, there appear gravitational descendant fields
$\sigma_n({\cal O}_\alpha)$ for $n=1,2, \cdots$. For the sake of
convenience we write $\Descd{0}{\alpha}=\Prim{\alpha}$. Note that the identity
operator ${\cal O}_0$ is identified with the puncture operator
$P$. Then, in the presence of gravity \metric\ should read
$$
\langle P {\cal O}_\alpha {\cal O}_\beta \rangle =\eta_{\alpha \beta}.
\eqno\eq
$$
In the case of pure gravity the only primary field is the puncture
operator $P$. The puncture operator, together with its first
descendant $\sigma_1(P)$
(dilaton field), plays a distinguished role in topological string
theory.

To each $\sigma_n({\cal O}_\alpha)$ we associate a coupling constant
$t_{n,\alpha}$. For short we write $t_\alpha=t_{0,\alpha}$ which are
called primary couplings henceforth. The coupling
constant space spanned by all $t_{n,\alpha}$ is referred to as the full phase
space, whereas its subspace with
$t_\alpha \not= 0, t_{n,\alpha}=0 ~(n \geq 1)$
is called the small phase space. When all $t_{n,\alpha}$ vanish the system
is on critical. Switching on $t_{n,\alpha}$ is equivalent to put the
system in non-trivial backgrounds. Evolution of the topological string
in such backgrounds is of our main interest.

We employ the flat coordinate system to describe the phase space [\KS,\DVV].
The flat coordinates are advantageous since in the small phase space
$\langle P {\cal O}_\alpha {\cal O}_\beta \rangle$ turn out to be independent
of $t_\alpha$ and generic three-point functions depending upon $t_\alpha$ are
evaluated from the free energy $F_0$
$$
C_{\alpha \beta \gamma}(t)={\partial^3 F_0 \over \partial t_\alpha
\partial t_\beta \partial t_\gamma} .
\eqno\eq
$$

The two-point functions
$$
u_{\alpha}=\langle P {\cal O}_{\alpha} \rangle
\eqno\eq
$$
will play a role of the fundamental order parameters in our topological string
theory. In the small phase space we have
$$
u_{\alpha}= \eta_{\alpha \beta} t_\beta .
\eqn\order
$$
This simple relation is invalidated in the full phase space where
$u_{\alpha}$ depend on $t_{n,\alpha}$ in a complicated way. In spite of
this fact there exists the following remarkable observation.
Suppose that a two-point function $\langle XY \rangle$ with
$X, Y \in \{\sigma_n({\cal O}_\alpha) \}$ is given as a function of
$t_\alpha$ in the small phase space. Note also that the relation \order\ is
invertible. Then we obtain $\langle XY \rangle$ expressed in terms of
$u_\alpha$, i.e.
$$
\bra XY \ket =R_{XY}(u).
\eqn\consti
$$
It was shown in [\DW] that the function $R_{XY}(u)$ is universal
in the sense that \consti\ continues to be valid even
in the full phase space. The relation \consti\ is thus quite fundamental,
and called the constitutive relation.
The proof is  based
on the topological recursion relation
$$
\langle \sigma_n({\cal O}_\alpha) XY \rangle
= \langle \sigma_{n-1}({\cal O}_\alpha) {\cal O}_\beta \rangle
\langle {\cal O}^\beta XY \rangle
\eqn\recursion
$$
with ${\cal O}^\beta= \eta^{\beta \gamma}{\cal O}_\gamma$.
We note that our normalization of $t_{n,\alpha}$ is different from [\DW],
and hence there is no factor $n$ on the RHS of \recursion .

The other important recursion relation is the puncture equation [\W,\DW]
$$
\langle P\sigma_{n_1}({\cal O}_{\alpha_1}) \cdots
\sigma_{n_s}({\cal O}_{\alpha_s}) \rangle
=\sum_{i=1}^s \langle
\prod_{k=1}^s \sigma_{n_k-\delta_{ik}}({\cal O}_{\alpha_k}) \rangle
\eqn\puncture
$$
which is valid in the small phase space. Using the full phase space free
energy $F_0$ we can cast \puncture\ into the differential equation
$$
{\partial \over \partial t_0}F_0
={1 \over 2} t_\alpha \eta_{\alpha\beta}t_\beta
+\sum_{n=0}^\infty  t_{n+1,\beta}
{\partial \over \partial t_{n,\beta}} F_0.
\eqno\eq
$$
Taking a derivative with respect to $t_\alpha$ one finds
$$
u_\alpha=\eta_{\alpha\beta}t_\beta
+\sum_{n=0}^\infty  t_{n+1,\beta}
\langle \sigma_n({\cal O}_\beta){\cal O}_\alpha \rangle ,
\eqn\stringeq
$$
where
$$
u_\alpha={\partial^2 F_0 \over \partial t_0 \partial t_\alpha},
\hskip1cm
\langle \sigma_n({\cal O}_\beta){\cal O}_\alpha \rangle
={\partial^2 F_0 \over \partial t_{n,\beta} \partial t_\alpha}.
\eqno\eq
$$
This is the string equation at genus zero [\DW].
Since the RHS is written in terms
of $u_\alpha$ (constitutive relation),
\stringeq\ is a kind of mean field
theoretic equation to determine the order parameters $u_\alpha$.
The puncture equation \puncture\ in fact holds for arbitrary genus.
Thus we have the string equation for all genera to which we turn in later
sections.

The genus zero free energy is known to take the universal form which
is now described as

{\bf Proposition 1}.
The free energy at $g=0$ is given by
$$
F_0(t)={1 \over 2}\sum_{n,m=0}^\infty
\langle \sigma_n({\cal O}_\alpha)\sigma_m({\cal O}_\beta) \rangle
\tilde t_{n,\alpha}\tilde t_{m,\beta},
\eqn\dubro
$$
where  $\tilde t_{n,\alpha}
=t_{n,\alpha}-\delta_{n1}\delta_{\alpha 0}$ [\K,\TT,\D].

This formula is not that obvious since two-point functions have the
non-trivial $t_{n,\alpha}$ dependence.

[proof] It is sufficient if one can show
$$
\sum_{n,m=0}^\infty
\Big( {\partial \over \partial t_{k,\gamma}}
\langle \sigma_n({\cal O}_\alpha)\sigma_m({\cal O}_\beta) \rangle \Big)
\tilde t_{n,\alpha}\tilde t_{m,\beta}=0.
\eqn\sufficient
$$
 From the constitutive relation we can make use of the chain rule
$$
{\partial \over \partial  t_{k,\gamma}}
= {\partial u_\rho \over \partial  t_{k,\gamma}}
{\partial \over \partial u_\rho}.
\eqn\chainrule
$$
To calculate the $u_\rho$-derivative we first note that
$$
\eqalign{
{\partial \over \partial t_\tau}
\langle \sigma_n({\cal O}_\alpha)\sigma_m({\cal O}_\beta) \rangle
& = \langle \sigma_n({\cal O}_\alpha)\sigma_m({\cal O}_\beta)
              \Prim{\tau}\rangle   \cr
& = \langle \sigma_{n-1}({\cal O}_\alpha){\cal O}_\sigma \rangle
\langle \sigma_{m-1}({\cal O}_\beta){\cal O}_\delta \rangle
\langle {\cal O}^\sigma {\cal O}^\delta \Prim{\tau} \rangle ,  \cr}
\eqno\eq
$$
where we have used \recursion\ twice. On the other hand the LHS of the
above may be evaluated by using the chain rule \chainrule. Thus we get
$$
{\partial \over \partial u_\rho}
\langle \sigma_n({\cal O}_\alpha)\sigma_m({\cal O}_\beta) \rangle
=\langle \sigma_{n-1}({\cal O}_\alpha){\cal O}_\sigma \rangle
\langle \sigma_{m-1}({\cal O}_\beta){\cal O}_\delta \rangle
C^{\sigma\delta\rho} ,
\eqn\uderivative
$$
where
$$
C^{\sigma\delta\rho}
=g^{\rho\tau} \bra {\cal O}^\sigma {\cal O}^\delta \Prim{\tau} \ket
\eqno\eq
$$
and $g_{\alpha\beta}g^{\beta\gamma}=\delta_\alpha^\gamma$ with
$g_{\alpha\beta}
=\partial^3 F_0 /\partial t_0\partial t_\alpha \partial t_\beta$.
Plugging \uderivative\ into the LHS of
\sufficient\ and using the string equation
one achieves the desired result.[\hskip -0.2mm]

In a mean field theory the variation of the free energy with respect to the
order parameters gives rise to the self-consistent mean field equation.
What do we obtain if we take a  variation of the free energy \dubro\
with respect to $u_\alpha$? To see this define
$u^\rho= \eta^{\rho\gamma}u_\gamma$ and evaluate
$\partial F_0 /\partial u^\rho$. Using \uderivative\ we get
$$
{\partial F_0 \over \partial u^\rho}
= {1 \over 2} C_\rho^{~\alpha\beta}(u) X_\alpha X_\beta
\eqno\eq
$$
with
$$
X_\alpha
=\Big( -u_\alpha+ \eta_{\alpha\gamma}t_\gamma
+\sum_{n=0}^\infty  t_{n+1,\gamma}
\langle \sigma_n({\cal O}_\gamma){\cal O}_\alpha \rangle \Big).
\eqno\eq
$$
Thus under the extremum condition $\partial F_0 /\partial u^\rho=0$ we obtain
$(\hbox{string equation})^2$=0. For pure gravity this was noted earlier
in [\IZ,\Y].

Let us next turn to explicit calculations of genus zero correlation functions.
We first prove that

{\bf Proposition 2}. The quantities
$$
u_{\alpha_1 \cdots \alpha_n} \equiv {\partial^{n+1} F_0 \over \partial t_0
\partial t_{\alpha_1} \cdots \partial t_{\alpha_n}}
\eqno\eq
$$
can be expressed as a sum of tree graphs with $n$-external lines.
The propagator and vertices are given in the proof.

[proof]
The genus zero string equation \stringeq\ is written as
$$
\eqalign{
&u_{\alpha}= \eta_{\alpha \beta} t_{\beta}+f_{\alpha}(t,u),\cr
&f_{\alpha}(t,u)=\sum_{n=0}^{\infty}
t_{n+1,\beta} \langle \sigma_{n}({\cal O}_{\beta}) {\cal O}_{\alpha}
\rangle.\cr
}
\eqno\eq
$$
Taking the derivatives of this equation with respect to
$t_{\beta}$ and using the fact that $f_{\alpha}$
depend on primary couplings only through the order parameters, one obtains
$$
\eqalign{
&u_{\alpha \beta}=\eta_{\alpha \beta}+f_{\alpha \gamma} u^{\gamma}_{\beta}, \cr
&u_{\alpha \beta \gamma}=
f_{\mu \nu \rho} u^{\mu}_{\alpha} u^{\nu}_{\beta} u^{\rho}_{\gamma}, \cr
&u_{\alpha \beta \gamma \delta}=
f_{\mu \nu \rho \sigma} u^{\mu}_{\alpha} u^{\nu}_{\beta} u^{\rho}_{\gamma}
u^{\sigma}_{\delta}+[f_{\mu \nu \lambda} u^{\lambda \tau} f_{\tau \rho \sigma}
u^{\mu}_{\alpha} u^{\nu}_{\beta} u^{\rho}_{\gamma} u^{\sigma}_{\delta}
+(2 \  {\rm more} \ {\rm terms})], \cr
&\cdots \ ,
}
\eqn\tree
$$
where
$$
f_{\alpha \alpha_1 \cdots \alpha_n}={\partial^n f_{\alpha} \over
\partial u^{\alpha_1} \cdots \partial u^{\alpha_n}}.
\eqn\fderivative
$$
We have derived the second line in \tree\ by noticing that the first line
is written as
$$
\delta_\alpha^\rho -f_{\alpha\gamma} \eta^{\gamma\rho}
=[u^{-1}]_\alpha^\rho ,
\eqno\eq
$$
where $u_{\alpha\beta}[u^{-1}]^{\beta\gamma}=\delta_\alpha^\gamma$.
The rest of the formulas is obtained in a similar manner.
\tree\ are the desired expressions for $u_{\alpha_1 \cdots \alpha_n}$
where $u_{\alpha \beta}$ is the propagator
and \fderivative\ are the vertices
(the indices of $u$ are raised and lowered by
the metric $\eta$).[\hskip -0.2mm]

Since the vertices are easily obtained by using
the Gauss-Manin relation as we shall see below,
\tree\ provides an efficient way of
calculating the genus zero correlation functions.
Let us consider the $A_1$ model and the $A_2$ model as examples.

\vskip5mm
\undertext{The $A_1$ Model (Pure Topological Gravity)}

The $g=0$ string equation is simply given by
$$
u(t)=\sum_{n=0}^{\infty} t_n {u(t)^n \over n!},
\eqno\eq
$$
where we have set $t_n \equiv t_{0,n}$. Along the line explained
above one evaluates $u^{(n)}={\partial^n \over \partial t_0^n} u$
as follows
$$
\eqalign{
u'&={1\over M},
\cr
u''&={{G[2]}\over {{M^3}}},
\cr
u'''&={{3 {{G[2]}^2}}\over {{M^5}}} + {{G[3]}\over {{M^4}}},
\cr
u^{(4)}&={{15 {{G[2]}^3}}\over {{M^7}}} + {{10 G[2] G[3]}\over {{M^6}}} +
 {{G[4]}\over {{M^5}}},
\cr
u^{(5)}&={{105 {{G[2]}^4}}\over {{M^9}}} +
    {{105 {{G[2]}^2} G[3]}\over {{M^8}}} +
    {{10 {{G[3]}^2}}\over {{M^7}}} +
    {{15 G[2] G[4]}\over {{M^7}}} +
    {{G[5]}\over {{M^6}}},
\cr}
\eqn\uGrelation
$$
where $M=1-G[1]$ and
$$
G[k]=\sum_{n=0}^{\infty} t_{n+k} {u(t)^n \over n!}.
\eqno\eq
$$
The variables $G[n]$'s were first introduced in [\IZ].

\vskip5mm
\undertext{The $A_2$ Model}

In the $A_2$ model we have two primaries, $\Prim{0}=P$ and $\Prim{1}=Q$.
Let us denote $\bra PP \ket =u$ and $\bra PQ \ket =v$.
Following the $A_1$ case we
introduce the $G$-type variables as
$$\eqalign{
&G[k]=t_{k,1}+t_{k+1,0} u+t_{k+1,1} v+t_{k+2,0} uv+
t_{k+2,1} \Big( {u^3 \over 6}+{v^2 \over 2}\Big) +\cdots, \cr
&H[k]=t_{k,0}+t_{k+1,0} v+t_{k+1,1} {u^2 \over 2}+
t_{k+2,0} \Big( {u^3 \over 3}+{v^2 \over 2}\Big)
+ t_{k+2,1} {u^2v \over 2}+\cdots.
}
\eqno\eq
$$
These variables are characterized by the relation
$$
{\partial \over \partial v}
\pmatrix{G[n] \cr H[n]}=
\pmatrix{G[n+1] \cr H[n+1]},
\hskip1cm
{\partial \over \partial u}
\pmatrix{G[n] \cr H[n]}=
\pmatrix{H[n+1] \cr u G[n+1]}.
\eqno\eq
$$
The string equation is then written as
$$
u=G[0], \hskip15mm v=H[0].
\eqno\eq
$$
The propagators, the three- and four-vertices given in \tree\ and
\fderivative\ turn out to be
$$
\pmatrix{u_{00} &u_{01} \cr
         u_{10} &u_{11} \cr}=
{1 \over (1-H[1])^2-u G[1]^2}
\pmatrix{G[1] &1-H[1] \cr
         1-H[1] &u G[1] \cr},
\eqno\eq
$$
$$\eqalign{
&f_{000}=G[2],\ f_{001}=H[2],\ f_{011}=u G[2],\ f_{111}=G[1]+u H[2],\cr
&f_{0000}=G[3],\ f_{0001}=H[3],\ f_{0011}=u G[3],\ f_{0111}=G[2]+u H[3], \cr
&f_{1111}=2 H[2]+u^2 G[3].
}
\eqno\eq
$$

\vskip5mm
\undertext{General Case}

Let us define the $G$-type variables as
$$
G_{\alpha}[k]= \eta_{\alpha\beta} t_{k,\beta}+
\sum_{n=0}^{\infty}  t_{n+1+k,\beta}
\langle \sigma_{n}({\cal O}_{\beta}) {\cal O}_{\alpha} \rangle.
\eqn\Gvariable
$$
Then the string equation \stringeq\ becomes
$$
u_{\alpha}=G_{\alpha}[0].
\eqn\Gstringeq
$$
Note that $G_{\alpha}[n]$ is characterized by the Gauss-Manin relation
$$
{\partial \over \partial u^{\beta}} G_{\alpha}[n]=
C_{\alpha \beta}^{~~~\gamma}(u) G_{\gamma}[n+1],
\eqno\eq
$$
which can be derived from the constitutive relation and \recursion.

The relation between $F_0$ and tree graphs is seen from another
point of view. From the free energy in the form \dubro\ one has
$$
{\partial F_0 \over \partial t_0}=
\sum_{m=0}^{\infty} \langle P \sigma_m ({\cal O}_{\beta}) \rangle
{\tilde t_{m,\beta}}.
\eqno\eq
$$
Considering this  as a function of $t$ and $u$, say $S(t,u)$,
it is easy to show that
$$
{\partial S \over \partial u_{\alpha}}=0
{}~\Longleftrightarrow
{}~{\rm string \ equation},
\eqn\variation
$$
which may be regarded as the action principle (at $g=0$)  [\Gins,\JY,\Wi].
Moreover, having the solution $u(t)$ of \variation\ it follows that
$$
S(t,u(t))={\partial F_0 \over \partial t_0}.
\eqno\eq
$$
Hence, ${\partial F_0 /  \partial t_0}$ can be evaluated
by the tree level Feynman diagrams with the action $S$.

In the $A_1$ and $A_2$ cases it turns out that the explicit form of
the action is very simple. We find
$$
\eqalign{
S_{A_1}&=-{1 \over 2}u^2+\sum_{n=0}^{\infty} t_n {u^{n+1} \over (n+1)!}, \cr
S_{A_2}&=-u v+\sum_{n,m=0}^{\infty} (m-2)!! t_{3n+2m-1}
 {v^{n} \over n!}{u^{m} \over m!}, \cr
}
\eqn\Atwoaction
$$
where, in $S_{A_2}$,
$t_n$ ($n \equiv \hskip -4mm / \hskip 3mm 0$ mod $3$) is defined by
$t_{3 n+\alpha+1}=t_{n, \alpha}$ and
$t_n=0$ if $n \equiv 0$ mod $3$ or $n < 0$.

Finally we remark that
the following observations will be important in subsequent sections.

{\bf Lemma 1}.
The relations, \Gvariable\ and \Gstringeq,
defining the transformation from the couplings $t_{n, \alpha}$
to the moments $G_{\alpha}[n]$ are invertible as
formal power series, that is
$$
t_{n,\alpha} \in {\bf Q}[[G_{\beta}[m] \  |\  \beta \in I; m \geq 0]].
\eqno\eq
$$
where ${\bf Q}[[x_1,x_2,\cdots]]$
means the ring of the formal power series
of $x_1,x_2,\cdots$ with coefficients in ${\bf Q}$.

[proof]
\Gvariable\ says
$$
t_{n, \alpha}=\eta_{\alpha \beta}
\Big\{ G_{\beta}[n]-\sum_{m=n+1}^{\infty} t_{m, \beta}
\Big[ {\rm polynomial \ in \ } u_{\gamma} \big(=G_{\gamma}[0]\big) \Big]
\Big\}.
\eqn\invGvariable
$$
These relations are of triangular form in $n$, and hence
by iterative substitution of these, one obtains the results.[\hskip -0.2mm]

Notice that, by solving \uderivative\ we
can express $f_{\alpha_1 \cdots \alpha_n}$'s (and hence $G_{\alpha}[n]$'s)
as polynomials in $u_{\alpha}$, $u_{\alpha \beta}$, $[u^{-1}]_{\alpha \beta}$,
$u_{\alpha \beta \gamma}$,
$\cdots$. Thus we get

{\bf Lemma 2}.
$$
t_{n,\alpha}
\in {\bf Q}[[u_{\alpha}, u_{\alpha \beta}, [u^{-1}]_{\alpha \beta},
u_{\alpha \beta \gamma}, \cdots
\  |\  \alpha, \beta \cdots \in I]].
\eqno\eq
$$

\chapter{Topological Strings at Higher Genera}

Let $F_g$ be the genus $g$ free energy of topological gravity
coupled to topological matter
associated with a simply laced Lie algebra $\cal G$. The genus expansion
of the free energy then reads $\log Z=F=\sum_{g=0}^{\infty} \lambda^{2g-2}F_g$
with $\lambda$ being the string coupling constant. The partition function
$Z$ is characterized by the following conditions [\FKN,\DVVer]:

{\bf Condition 1}: $Z$ is a $\tau$-function of the Drinfeld-Sokolov hierarchies
associated to the Lie algebra ${\cal G}$.

{\bf Condition 2}: $Z$ satisfies the $W$-constraints
associated with the
Lie algebra ${\cal G}$ (twisted by the Coxeter element):
$$
W^{(s+1)}_{n} Z=0, \ (s \in I, \ n \geq -s).
\eqn\wconstraint
$$
Among the $W$-constraints the $L_{-1}$ condition is equivalent to the
puncture equation
$$
{\partial F_g\over \partial t_{0,0}}=
{1 \over 2} \eta_{\alpha \beta} t_{0,\alpha} t_{0, \beta} \delta_{g,0}+
\sum_{n=1}^{\infty} t_{n+1, \beta}
{\partial F_g \over \partial t_{n, \beta}}.
\eqn\punctureeq
$$
The $L_0$ condition reduces to the dilaton equation
$$
{\partial F_g \over \partial t_{1,0}}=
(2g-2) F_g+
\sum_{n=0}^{\infty} t_{n,\beta}
{\partial F_g \over \partial t_{n, \beta}}.
\eqn\dilatoneq
$$
The conditions 1 and 2 are highly over determined system
and many of them are redundant. In fact, in the case of $A_1$, $A_2$
models, one can prove that

{\bf Lemma 3}. $F_g$ can be determined uniquely by
the puncture equation, the dilaton equation
and the flow equations with respect to the primary
couplings $\{ t_{0,\alpha} \}$ and the dilaton coupling $t_{1,0}$.

[proof] Consider a correlation function of primary operators
$$
\langle O_{\alpha_1} \cdots O_{\alpha_N} \rangle_g.
\eqno\eq
$$
When all the primary couplings are taken to be zero, these correlators
are sum of finite terms of the form
$$
(1-t_{1,0})^{-(N+\sum_{(n,\alpha) \in I_+} k_{n,\alpha}+2g-2)}
\prod_{(n,\alpha) \in I_+} t_{n,\alpha}^{k_{n,\alpha}},
\eqno\eq
$$
where $I_+=\{(n,\alpha) \vert ~(n,\alpha) \neq (1,0), (0,\alpha) \}$.
The coefficients of these terms can be determined by using
the primary and dilaton flow equations, and the puncture equation via
induction on $g$ and $N$.
[\hskip -0.2mm]

It seems likely that
this uniqueness lemma also holds for all ADE-type theories. In the following
we derive some explicit formulas for $F_g$ in the $A_1$ and $A_2$ cases
making use of lemma 3.

\section{The $A_1$ model}

The flow in the $A_1$ model is governed by the KdV equation. Some data on
the $A_1$ potentials are available in appendix A.
By simple degree counting, $F_g$ is homogeneous of degree $3(1-g)$
under the assignment
${\rm deg}(t_i)=1-i$.
The first few terms of $F_g$ are
$$
\eqalign{
&F_0={t_0^3 \over 3!}+t_1 {t_0^3 \over 3!}+t_2 {t_0^4 \over 4!}+
    2 {t_1^2 \over 2!} {t_0^3 \over 3!}+\cdots, \cr
&F_1={1 \over 24} \Big( t_1+ {t_1^2 \over 2!}+t_0 t_2+\cdots \Big), \cr
&F_2={1 \over 5760} (5 t_4+\cdots).
}
\eqno\eq
$$

For $g=1$ the closed expression for $F_1$ has been obtained [\DW,\IZ,\Amb].
The result is
$$
F_1={1 \over 24} \log u',
\eqno\eq
$$
where $u=F''_0$. For $g \geq 2$ one can prove that

{\bf Theorem 1}: There exists a formula for $F_g$ ($g>1$)
of the form:
$$
\eqalign{
F_g
=&\sum_{\sum (k-1) l_k=3g-3} a_{l_2 \cdots l_{3g-2}}
{[u'']^{l_2} \cdots [u^{(3g-2)}]^{l_{3g-2}} \over
[u']^{2(1-g)+\sum k l_k}} \cr
=&\sum_{\sum (k-1) l_k=3g-3} b_{l_2 \cdots l_{3g-2}}
{[G[2]]^{l_2} \cdots [G[3g-2]]^{l_{3g-2}} \over
[u']^{2(1-g)+\sum l_k}}.
}
\eqn\formulafg
$$

[proof]
We will prove the first line of the formula.
The second one is obtained with the aid of \uGrelation.
As we have shown in the previous section, the coupling $t_i$ is solved
as a formal power series in $u,u',{1 \over u'},u'',u''',\cdots$.
Then one has an expression of the form
$$
F_g \in {\bf Q}[[u,u',{1 \over u'},u'',u''',\cdots]].
\eqno\eq
$$
We rewrite the puncture equation \punctureeq\ as
$$
{\cal L}_{-1} F_g \equiv
\Big( {\partial \over \partial t_0}-
\sum_{n=0}^{\infty} t_{n+1} {\partial \over \partial t_n} \Big) F_g
=\delta_{g,0} {t_0^2 \over 2}.
\eqno\eq
$$
Especially for $g=0$ one obtains
$$
{\cal L}_{-1} u^{(k)}=\delta_{k,0}.
\eqno\eq
$$
These two equations imply that $F_g$ ($g > 0$) cannot depend on $u$.
Since there is no variable of positive degree other than $u$,
the number of available variables is finite. The highest derivative
variable is $u^{(3g-2)}$.
The power of degree zero factor $u'$
is fixed since each term of $F_g$ should have
a total number of $(2g-2)$ $t_0$-derivatives.
This proves the theorem.[\hskip -0.2mm]

Thus we are left with a finite number of unknown coefficients at each genus.
These coefficients can be fixed by using the first nontrivial
constitutive relation
$$
\langle P \sigma_1 \rangle={u^2 \over 2}+{u'' \over 12}.
\eqno\eq
$$

We mention here that
each term in $F_g$ can be associated with a
$g$-loop graph with $g-1$ external lines, in which $1/u'$ and $u^{(n+1)}$ play
the role of the propagator and $n$-vertex. This correspondence was first
noticed in [\DW]. The second expression in
\formulafg\ was obtained earlier in [\IZ].
Furthermore they point
out that the coefficients $b_{l_2 \cdots l_{3g-2}}$ yield the intersection
numbers on moduli space of Riemann surfaces.

In appendix D we will list the explicit formulas up to $g=4$.
For $g \geq 2$ the higher derivative terms in $F_g$ are given by
$$
F_g=a_{1} {[u^{(3g-2)}] \over [u']^g}+
    a_{2} {[u''] [u^{(3g-3)}] \over [u']^{g+1}}+
    a_{3} {[u'''] [u^{(3g-4)}] \over [u']^{g+1}}+
    a_{4} {[u'']^2 [u^{(3g-4)}] \over [u']^{g+2}}+\cdots,
\eqno\eq
$$
where
$$
\eqalign{
a_{1}&={1 \over 24^g g!}, \cr
a_{2}&={-21 \over 24^g (g-2)!}, \cr
a_{3}&={3(65-57g) \over 70 24^g (g-2)!}, \cr
a_{4}&={-3(1226+325g-1029g^2) \over 1440 24^g (g-2)!}.
}
\eqno\eq
$$
The coefficient $a_1$ has already been obtained in [\IZ].

\section{The $A_2$ model}

In the $A_2$ case,
the degrees of couplings and the free energy are
deg$(t_{n,\alpha})=3(1-n)-\alpha$, and deg$(F_g)=8(1-g)$.
Constitutive relations and two-point functions are given explicitly in
appendix B.
The puncture and dilaton equations for the $A_2$ model read
$$
\eqalign{
&{\partial F_g \over \partial t_{0,0}}=
t_{0,0} t_{0,1} \delta_{g,0}+
\sum_{n=0}^{\infty} t_{n+1,\alpha}
{\partial F_g \over \partial t_{n,\alpha}}, \cr
&{\partial F_g \over \partial t_{1,0}}=
(2g-2) F_g+
\sum_{n=0}^{\infty} t_{n,\alpha}
{\partial F_g \over \partial t_{n,\alpha}}.
}
\eqno\eq
$$
The flow equations can be read off from the constitutive relations in
appendix B.

Now, we study the generalization of the $A_1$ result (Theorem 1).
Let us write the genus zero variables as
$$
u={\partial^2 F_0 \over \partial t_{0,0} \partial t_{0,0} }, \hskip10mm
v={\partial^2 F_0 \over \partial t_{0,0} \partial t_{0,1} }.
\eqno\eq
$$
By the same argument as in the previous $A_1$ case,
one has an expansion of the form
$$
F_g \in {\bf Q}[[{1 \over u u'^2-v'^2}, u,v,u',v',u'',v'',\cdots]], \ (g>1).
\eqno\eq
$$
The degrees are deg$(u^{(n)})=2-3n$, deg$(v^{(n)})=3-3n$.
It is proved that $v$ can be eliminated from the expansion while
$u$ still survives. Since
$u$ has positive degree the expansion of $F_g$ will be an
infinite series.

In spite of this difficulty we can show that

{\bf Theorem 2}:
$$
\eqalign{
F_1&={1 \over 24} \log \det  u_{\alpha \beta}
={1 \over 24} {\rm log}(u u'^2-v'^2),\cr
F_2&={1 \over 1152} Q_1-{1 \over 360} Q_2-{1 \over 1152} Q_3
     +{1 \over 360} Q_4,\cr
}\eqno\eq
$$
where we have used $\bra PQQ \ket_0 =uu'$ (see appendix B) and
$$
\eqalign{
&Q_1=u_{0 \alpha \beta \gamma \delta} [u^{-1}]^{\alpha \beta}
[u^{-1}]^{\gamma \delta},\cr
&Q_2=u_{0 \alpha \beta \gamma} u_{\delta \mu \nu} [u^{-1}]^{\alpha \delta}
[u^{-1}]^{\beta \mu} [u^{-1}]^{\gamma \nu},\cr
&Q_3=u_{0 \alpha \beta} u_{\gamma \delta \mu \nu} [u^{-1}]^{\alpha \gamma}
[u^{-1}]^{\beta \delta} [u^{-1}]^{\mu \nu},\cr
&Q_4=u_{0 \alpha \beta} u_{\gamma \delta \mu} u_{\nu \rho \tau}
[u^{-1}]^{\alpha \gamma} [u^{-1}]^{\beta \nu} [u^{-1}]^{\delta \rho }
[u^{-1}]^{\mu \tau}.
}
\eqno\eq
$$

[proof] By construction, these formulas satisfy the puncture and
dilaton equations. The remaining task is to check the flow equations
(up to dilaton), which is straightforward (by using the computer).
[\hskip -0.2mm]

These results suggest that for the $A_2$ model

{\bf Ansatz}: The genus $g>0$ free energy $F_g$ can be expressed
as a sum of diagrams with $g$-loops and ($g-1$)-external lines with the
puncture index 0, in
which the  inverse propagator is $u_{\alpha \beta}$ and the $n$-vertex
$u_{\alpha_1 \cdots \alpha_n}$ is given by
$$
u_{\alpha_1 \cdots \alpha_n}={\partial^{n+1} F_0
\over \partial t_0 \partial t_{\alpha_1} \cdots \partial t_{\alpha_n}}.
\eqn\nvertex
$$

We have partial results for $g=3$
(there exist about one hundred terms),
$$
\eqalign{
F_3&={1 \over 82944} P_1
    -\Big[{1 \over 27648} P_2 +{1 \over 8640} P_3 \Big] \cr
   &-\Big[{1 \over 27648} P_4 +{29 \over 16128} P_5 +
      {1 \over 82944} P_6 +{73 \over 725760} P_7 \Big] +\cdots, \cr
}
\eqno\eq
$$
where $P_i$ are given by
$$
\eqalign{
P_1&=u_{00\alpha\beta\gamma\delta\mu\nu}~
[u^{-1}]^{\alpha\beta} [u^{-1}]^{\gamma\delta} [u^{-1}]^{\mu\nu},\cr
P_2&=u_{0\alpha\beta\gamma\delta\mu\nu}~ u_{0\rho\sigma}~
[u^{-1}]^{\alpha\rho} [u^{-1}]^{\beta\sigma}
[u^{-1}]^{\gamma\delta} [u^{-1}]^{\mu\nu}, \cr
P_3&=u_{00\alpha\beta\gamma\delta\mu}~ u_{\nu\rho\sigma}~
[u^{-1}]^{\alpha\beta} [u^{-1}]^{\gamma\nu}
[u^{-1}]^{\delta\rho} [u^{-1}]^{\mu\sigma}, \cr
P_4&=u_{00\alpha\beta\gamma\delta}~ u_{\mu\nu\rho\sigma} ~
[u^{-1}]^{\alpha\beta} [u^{-1}]^{\gamma\mu}
[u^{-1}]^{\delta\nu} [u^{-1}]^{\rho\sigma}, \cr
P_5&=u_{0\alpha\beta\gamma\delta\mu}~ u_{0\nu\rho\sigma}~
[u^{-1}]^{\alpha\beta} [u^{-1}]^{\gamma\nu}
[u^{-1}]^{\delta\rho} [u^{-1}]^{\mu\sigma}, \cr
P_6&=u_{00\alpha\beta}~ u_{\gamma\delta\mu\nu\rho\sigma}~
[u^{-1}]^{\alpha\gamma} [u^{-1}]^{\beta\delta}
[u^{-1}]^{\mu\nu} [u^{-1}]^{\rho\sigma}, \cr
P_7&=u_{00\alpha\beta\gamma\delta}~ u_{\mu\nu\rho\sigma}~
[u^{-1}]^{\alpha\mu} [u^{-1}]^{\beta\nu}
[u^{-1}]^{\gamma\rho} [u^{-1}]^{\delta\sigma}. \cr
}
\eqno\eq
$$

Furthermore, for any $g>1$, we can show that the highest derivative term
becomes
$$
F_g={1 \over 24^g g!} u_{_{\underbrace{00...0}_{g}} \alpha_1 \beta_1 \cdots
\alpha_g \beta_g}
[u^{-1}]^{\alpha_1 \beta_1} \cdots [u^{-1}]^{\alpha_g \beta_g}+\cdots.
\eqno\eq
$$
It is interesting to note that
all these diagrammatic formulas for $A_2$ coincide with that of $A_1$.
Then we have made some computations to see if the simple diagrammatic
structure observed for the $A_1,~A_2$ models exists also for other models.
According to our preliminary results on the $A_3$ and $D_4$ models
more elaborate structure seems to be required to represent the higher
genus terms.

\chapter{General Models}

In this section we wish to examine to what extent the observation made so far
is valid for more general models.
In [\W,\DW,\Wi] the general structure of
topological gravity coupled to topological matter is described.
The main result there is the ``topological recursion relation''
at $g=0$ and $g=1$. They read
$$
\langle \sigma_n({\cal O}_{\alpha}) X Y \rangle_0=
\langle \sigma_{n-1}({\cal O}_{\alpha}) {\cal O}_{\beta} \rangle_0
\langle {\cal O}^{\beta} X Y \rangle_0
\eqno\eq
$$
and
$$
\langle \sigma_n({\cal O}_{\alpha}) \rangle_1=
{1 \over 24} \langle \sigma_{n-1}({\cal O}_{\alpha})
{\cal O}_{\beta} {\cal O}^{\beta} \rangle_0 +
\langle \sigma_{n-1}({\cal O}_{\alpha}) {\cal O}^{\beta} \rangle_0
\langle {\cal O}_{\beta} \rangle_1 .
\eqno\eq
$$
The $g=1$ formula \genusone\ is
the direct consequence
of these equations [\DW,\Wi].
Hence, the problem of checking the validity of the $g=1$ formula reduces to the
consistency check between the flow equations and the above recursion relations.
This has been done at least for $A_p$ models
using the direct relation between the KP and
dispersionless KP formalism [\DW,\KNNT].

Recently Dubrovin has constructed
topological matter theory associated to each Coxeter group
(not only ADE but also BC and others) based on the WDVV equations [\Dub].
The existence as well as the uniqueness
of higher genus extension of these models is an interesting open question
to which we now address ourselves.

\section{Formulation of the problem}

To clarify the points we first state the problem clearly.
Let $F_0$ be any solution of the WDVV equation [\W,\DVV]\
$$
C_{\alpha \beta \mu} \eta^{\mu \nu} C_{\nu \gamma \delta}=
C_{\alpha \gamma \mu} \eta^{\mu \nu} C_{\nu \beta \delta},
\eqn\wdvveq
$$
where the notation is the same as in sect.2.
The flow equations consistent with \wdvveq\ should take the form
$$
{\partial u_{\beta} \over \partial t_\alpha}
=\big[ \bra \Prim\alpha \Prim\beta \ket_0 +
\lambda^2\; ({\rm terms} \ {\rm with} \ {\rm two} \ {\rm derivatives})
+O(\lambda^4) \big] '.
\eqno\eq
$$
where $'=\partial /\partial t_0$, $\lambda$ is a parameter and
terms of order $\lambda^{2n}$ consist of
differential polynomials in
$u$'s with a total number of $2n$ derivatives and suitable degree,
whose coefficients will be determined by the commutativity of the flows.
We now ask if such flows exist. If so, are they unique ?

For the ADE models it is believed that the answer is affirmative
and the unique flow corresponds to the  Drinfeld-Sokolov hierarchy.
We will check this in the $D_4$ case explicitly.

There are four primaries $P, Q, R$ and $S$ among which $S$ has the
${\bf Z}_2$ parity odd and the others are even.
The order parameters are
$$
   u=\langle {P} {P} \rangle,
 \ v=\langle {P} {Q} \rangle,
 \ w=\langle {P} {R} \rangle,
 \ f=\langle {P} {S} \rangle.
\eqno\eq
$$
The flow equations (or the constitutive relation) are determined by their
commutativity.  We start with the dispersionless ($\lambda=0$) terms
which are provided from the Landau-Ginzburg description using the
flat coordinate [\DVV].
It is remarkable that the commutativity condition determines
the full flows completely up to one parameter $\lambda$ which is
identified as the string coupling constant.
In appendix C we list the results of full genus flow up to the dilaton flow
equation.
We also checked that the $g=1$ terms are consistent with \genusone.

\section{Two-primary models}

To get some insight into the non-simply laced cases,
let us consider a two-primary model whose $g=0$ correlation functions are
defined by
$$
\langle PPQ \rangle_0=\langle Q^5 \rangle_0=1.
\eqn\generalmodel
$$
The flow equations consistent with the topological
recursion relation should be
$$
\eqalign{
{\partial u \over \partial t_{0,1}}&=[v]',\cr
{\partial v \over \partial t_{0,1}}&=
\Big[ {u^3 \over 6} + \lambda^2 \Big( {{{{u'}^2}}\over {24}} +
      {{uu''}\over {12}} \Big)  + {a_1}{\lambda^4}u^{(4)} \Big]',\cr
{\partial u \over \partial t_{1,0}}&=
\Big[ uv + \lambda^2 {v'' \over 6} + {a_2}{\lambda^4}u^{(4)} \Big]',\cr
{\partial v \over \partial t_{1,0}}&=
\Big[ {u^4 \over 8} + {{{v^2}}\over 2} +
   \lambda^2 \Big( {{u{{u'}^2}}\over 6} + {{{u^2}u''}\over 6} \Big)  +
   {\lambda^4}\big( {a_5}{{u''}^2} + {a_4}u'u^{(3)} +
      {a_3}uu^{(4)} + {a_6}v^{(4)} \big) \Big]' ,\cr
}
\eqn\twoprimaryflow
$$
where $a_1,\cdots ,a_6$ are numerical coefficients yet to be determined.

In the RHS of \twoprimaryflow\ the $g=0$ terms are obtained from
the constitutive relations (see sect.2.3 in [\DW]).
The $g=1$ terms are determined by imposing
the mutual commutativity of flows. We find that the result is unique up to
an overall constant which is absorbed into $\lambda^2$.
This $g=1$ result is also consistent with the genus one
free energy \genusone, which is seen by expanding, for instance,
$u=\sum_{g=0}^\infty \lambda^{2g} u_g$ and evaluating
$\partial u_1 / \partial t_{1,0}=\partial^3 F_1 /\partial t_{1,0}
\partial t_{0,0}^2$ with the aid of the $g=0$ flow equations.

At $g=2$ the commutativity of the flows demands that
$$
\eqalign{
0=\Big[ {\partial \over \partial t_{0,1}},
{\partial  \over \partial t_{1,0}}\Big] u
&=\lambda^4 \Big(
( {7\over {72}} - {a_4} - 2{a_5} ) {{u^{(3)}}^2} +
   ( {{11}\over {72}} - {a_3} - 2{a_4} - 2{a_5} ) u''u^{(4)}
\cr &
    + ( {5\over {72}} + {a_1} - 2{a_3} -
      {a_4} ) u'u^{(5)}
\cr &
   +( {1\over {72}} + {a_1} - {a_3} ) uu^{(6)} +
   ( {a_2} - {a_6} ) v^{(6)} \Big) +O(\lambda^6),\cr
0=\Big[ {\partial \over \partial t_{0,1}},
{\partial  \over \partial t_{1,0}}\Big]
v &=\lambda^4 \Big(
15{a_6}{{u''}^3} + ( -{1\over {72}} - 20{a_1} + {a_4} +
      2{a_5} ) u^{(3)}v^{(3)} +
   15{a_6}{{u'}^2}u^{(4)}
\cr &
   +   ( -15{a_1} + {a_3} + {a_4} ) v''u^{(4)} +
   ( -{1\over {36}} - 15{a_1} + {a_4} + 2{a_5} ) u''v^{(4)}
\cr &
     + ( -5{a_1} + {a_3} ) v'u^{(5)} +
   u'( 60{a_6}u''u^{(3)} +
      ( -{1\over {36}} - 6{a_1} + {a_3} + {a_4} )
       v^{(5)} )
\cr &
    +  ( {{-{a_2}}\over 2} +
      {{{a_6}}\over 2} ) {u^2}u^{(6)} +
      u( 10{a_6}{{u^{(3)}}^2} + 15{a_6}u''u^{(4)} +
      ( -{a_2} + 6{a_6} ) u'u^{(5)}
\cr &
      +( -{1\over {72}} - {a_1} + {a_3} ) v^{(6)} ) \Big) +O(\lambda^6) .\cr
}
\eqno\eq
$$
Unfortunately these two equations are inconsistent (no solution exist for
$a_1, \cdots , a_6$). Thus the model \generalmodel\ cannot be extended to
higher genus beyond $g=1$.

In general, for a two-primary model having $g=0$ correlation functions
$$
\langle PPQ \rangle_0=\langle Q^{s+2} \rangle_0=1,
\eqno\eq
$$
the flow equations up to $g \leq 1$ take the form
$$
\eqalign{
{\partial \over \partial t_{0,1}} u &=[v]',\cr
{\partial \over \partial t_{0,1}} v &=
\Big[ {{{u^s}}\over s!} + \lambda^2 \Big( {u^{s-3} \over {(s-3)!}}
{{{{u'}^2}}\over {24}} +
      {u^{s-2} \over {(s-2)!}} {{u''}\over {12}} \Big) \Big]',\cr
{\partial \over \partial t_{1,0}} u
&=\Big[ uv + {{\lambda^2v''}\over 6} \Big] ',\cr
{\partial \over \partial t_{1,0}} v &
=\Big[ s {u^{s+1} \over {(s+1)!}} + {{{v^2}}\over 2} +
   \lambda^2(s+1) \Big( {u^{s-2} \over {(s-2)!}}{{{u'}^2}\over 24} +
   {u^{s-1} \over {(s-1)!}}{{u''}\over 12} \Big) \Big] ' .\cr
}
\eqno\eq
$$
These $g=1$ terms are again consistent with \genusone.
Except for the case of $s=2$ (the Boussinesq equations), however,
one cannot generalize these equations for $g>1$,
which we have checked explicitly for $s=3,4,5$.

The spectrum of this two-primary model is characterized by the
exponents $I=\{1,s\}$ and the Coxeter number $s+1$.
Hence the case $s=3$ (or $s=5$) corresponds to the $B_2(=C_2)$ model
(or $G_2$ model)
of the Dubrovin's table [\Dub].
Of course, there exist integrable hierarchies
corresponding to these algebras. The above results show that these hierarchies
are not consistent with the topological recursion relation.
This result could be related to the fact that one needs more than one
$\tau$-functions in the Hirota formalism of these hierarchies.
It is rather remarkable that the consistency of these models fails at the
two-loop level.

\section{THE $CP^1$ MODEL}

We now discuss the $CP^1$ topological sigma model coupled to gravity [\W].
The spectrum of the model consists of two primaries, $P$ and $Q$, and their
descendants. Classically $P$ and $Q$ correspond to the zero form and the
K\"ahler form
which generate the cohomology of $CP^1$. We shall see that the $CP^1$ flow
equations have higher genus extension.

The constitutive relation for $\bra QQ \ket$ is given by
$$
\bra QQ \ket = e^u
\eqno\eq
$$
with $u=\bra PP \ket$ and $v=\bra PQ \ket$ [\DW]. This characteristic form of
$\bra QQ \ket$ realizes the idea of a quantum deformation of the classical
cohomology ring due to the instanton effect [\W]. The other genus zero
functions $\bra \sigma_n(X)Y \ket$ are obtained using the topological
recursion relations [\DW].

To consider the flow equations in the $CP^1$ model let us concentrate on the
primary and dilaton flow equations. We start with the dispersionless flows
$$
\eqalign{
{\partial u \over \partial t_{0,Q}}&=[v]' ,  ~~~~~~~~~~~~
{\partial v \over \partial t_{0,Q}}=[e^u]',  \cr
{\partial u \over \partial t_{1,P}}&=[uv]',  ~~~ ~~~~~~~
{\partial v \over \partial t_{1,P}}=[(u-1)e^u]', \cr}
\eqno\eq
$$
where the relations $\bra \sigma_1(P)P \ket =uv$ and
$\bra \sigma_1(P)Q \ket = (u-1)e^u$ have been
utilized. As in the previous section we want to determine higher genus terms
by imposing the commutativity among flows. In contrast to the Dubrovin type
models, it is difficult to enumerate candidate higher terms in the $CP^1$ model
since $u$ and $v$ do not
possess definite degrees. However, it turns out  that we can still fix
the higher genus terms consistently. The result reads
$$
\eqalign{
 {\partial u \over \partial t_{0,Q}} =&  [v]',   \cr
 {\partial v \over \partial t_{0,Q}} =&
\big[e^u + \lambda^2 G_1 e^u+ \lambda^4 G_2 e^u +\lambda^6 G_3 e^u
+O(\lambda^8) \big]',   \cr
 {\partial u \over \partial t_{1,P}} =&
\big[ uv + \lambda^2 H_1 + \lambda^4 H_2 +\lambda^6 H_3
+O(\lambda^8) \big]' , \cr
 {\partial v \over \partial t_{1,P}} = &
\Big[(u-1)e^u + \half v^2 + \lambda^2 K_1 e^u+ \lambda^4 K_2 e^u
+\lambda^6 K_3 e^u +O(\lambda^8) \Big]',  \cr}
\eqno\eq
$$
where
$$
\eqalign{
G_1 =&  {1 \over 24} u'^2+ {1 \over 12}u'',   \cr
G_2 =&  {1 \over 360}u'''' + {1 \over 180}u'u'''
+ {7 \over 1440} u'^2u''+ {1 \over 1920} u'^4
+{1 \over 240 }u''^2 ,  \cr
G_3 =&  {1 \over 20160} u^{(6)}+{1 \over 6720} u^{(5)} u'
+{19 \over 60480} u^{(4)}u''+{1 \over 5040} u^{(4)} u'^2  \cr
&+{23 \over 120960} u'''^2+{41 \over 60480} u'''u''u'+{1 \over 6720} u'''u'^3
+{29 \over 181440} u''^3   \cr
&+{37 \over 120960} u''^2u'^2+{11 \over 161280} u''u'^4
+{1 \over 322560} u'^6 ,  \cr}
\eqno\eq
$$
$$
\eqalign{
H_1=& {1 \over 6} v'',  \cr
H_2 =& -{1 \over 360} v'''',  \cr
H_3=& {1 \over 15120} v^{(6)},  \cr}
\eqno\eq
$$
$$
\eqalign{
K_1=& {1 \over 24} (u+3)u'^2+{1 \over 12} (u+2)u'', \cr
K_2 =&  {1 \over 360}(u+4)u''''+{1 \over 180} (u+5)u'u'''
+{7 \over 1440} (u+6)u'^2u''   \cr
&+{1 \over 1920}(u+7)u'^4+{1 \over 240}(u+5)u''^2 ,   \cr
K_3=&  {1 \over 20160}(u+6) u^{(6)}+{1 \over 6720} (u+7)u^{(5)} u'
+{19 \over 60480}(u+7) u^{(4)}u''  \cr
&+{1 \over 5040}(u+8) u^{(4)} u'^2+{23 \over 120960} (u+7) u'''^2
+{41 \over 60480} (u+8) u'''u''u'  \cr
&+{1 \over 6720}(u+9) u'''u'^3
+{29 \over 181440} (u+8) u''^3 +{37 \over 120960}(u+9) u''^2u'^2  \cr
&+{11 \over 161280} (u+10)u''u'^4
+{1 \over 322560}(u+11) u'^6 .  \cr}
\eqno\eq
$$
 From the order $\lambda^2$ terms  we have confirmed that the genus
one free energy again takes the form of \genusone.

These computations show that the $CP^1$ model has in fact a consistent
higher genus extension which would provide valuable information on the
holomorphic maps from higher-genus Riemann surfaces onto $CP^1$. One of
us (T.E.) has recently obtained a preliminary result that the underlying
integrable hierarchy of the $CP^1$ model is the Toda lattice hierarchy
as suggested in [\D]
(with a constraint that two kinds of Toda times $\{ t_n\} ,\{ \bar t_n\}$
coincide $t_n=\bar t_n, (n=1,2,\cdots)$) and the times $t_{n,Q}$ associated
with the $Q$ operator are identified with the Toda times $t_{n,Q}=t_{n+1},
(n=0,1,2, \cdots)$. Details of the $CP^1$ model will be discussed
elsewhere.

\chapter{Discussions}

What we have been trying to do in the present paper is somewhat similar
to the perturbation series in quantum field theories. In the Feynman
diagram expansion of conventional field theories the amplitudes associated
with multi-loop diagrams are written in terms of the tree level propagators
and vertices according to the Feynman rule. In topological string theory
our intention has been to organize the genus expansion in such a way that
the $g \geq 1$ free energy at each genus is expressed in terms of the
genus zero quantities. We have partially succeeded in pursuing this
program through explicit but tedious calculations. The full
clarification remains still open.

In order to embody an idea it is instructive to examine
a simpler geometrical model. Taking the vector model as such an example
we present the calculations in a way parallel to those for gravity models.

\vskip5mm
\undertext{The Vector Model}

The O(N) vector model has been considered by several groups
[\NY,\DKO,\AMP] to
enhance our understanding of the matrix model for two-dimensional
gravity. We follow closely ref.[\NY] and perform the calculations in the light
of our spirit.

The O(N) model describes geometrical critical phenomena of randomly
branched polymers. In the double scaling limit the free energy $F$ is
defined from the partition function $Z(t)$ as
$F(t)=\log Z(t)$ where we turn on infinitely many coupling constants $t_n~
(n=0,1,2,\cdots)$ which are conjugate to the scaling operators $\sigma_n$.
In particular $\sigma_0$ is the marking operator $P$ in the theory.
There exists a ``polymer'' coupling constant $\lambda$ with respect to which
we are going to define the ``genus'' expansion
$F=\sum_{g=0}^\infty \lambda^{g-1} F_g$.

Introduce an one-point function $\bra P \ket =F'\equiv f/\lambda$
with $'=\partial / \partial t_0$ which is the basic order parameter,
and hence the counterpart of $u_\alpha$ in gravity theory.
The ``polymer'' equation, i.e. the analog of the string equation, now
reads [\NY]
$$
t_0+\sum_{n=1}^{\infty} t_n R_{n-1}=0,
\eqn\polymereq
$$
where $R_n/\lambda =\bra \sigma_n \ket =\partial F/\partial t_n$
(i.e. $R_0=f$). We demand that the flows parametrized by $t_n$ commute.
Then one has
$$
{\partial \over \partial t_n}f=[R_n]',
\eqn\polflow
$$
thereby $R_n$ become the flow potentials. These flow potentials obey
$$
(n+1) R_n = \lambda [R_{n-1}]'+f R_{n-1}.
\eqn\polrecursion
$$
$R_n$ has an expansion of the form
$$
R_n={f^{n+1} \over (n+1)!}+
    \lambda {1 \over 2} {f^{n-1}\over (n-1)!} f'+
    \lambda^2 \Big[ {1 \over 8} {f^{n-3}\over (n-3)!} f'^2+
    {1 \over 6} {f^{n-2}\over (n-2)!} f'' \Big]+\cdots.
\eqno\eq
$$
Explicitly one finds
$$
\eqalign{
R_1&={1 \over 2} (f^2+\lambda f'), \cr
R_2&={1 \over 6} (f^3+3 \lambda f f'+\lambda^2 f''), \cr
R_3&={1 \over 24} (f^4+6 \lambda f^2 f'
+4\lambda^2 f f''+3\lambda^2 f'^2+\lambda^3 f''').
}
\eqn\polpotentials
$$

Let us expand $f=\sum_{g=0}^\infty \lambda^g f_g$,
then it is convenient to introduce the moments $I_k$ as
$$
I_k=\sum_{n=0}^{\infty} t_{n+k} {f_0^n \over n!}.
\eqno\eq
$$
The following relations are useful
$$
f_0'=-{1 \over I_1}, \hskip10mm
I_n'=-{I_{n+1} \over I_1}.
\eqn\momrelation
$$

It is now straightforward to evaluate $f_g$ using $I_k$.
Having \polpotentials\ we first expand
the polymer equation as follows
$$
\eqalign{
0=&I_0+\lambda \Big( I_1 f_1+{1 \over 2} f_0'\Big) \cr
&+\lambda^2 \Big( I_1 f_2+{1 \over 2} I_2 f_1^2+{1 \over 2} I_3 f_1 f_0'+
 {1 \over 2} I_2 f_1'+{1 \over 8} I_4 f_0'^2+{1 \over 6} I_3 f_0'' \Big)
+\cdots }
\eqno\eq
$$
>from which one obtains
$$
\eqalign{
0&=I_0=t_0+\sum_{n=1}^{\infty} t_{n} {f_0^n \over n!},\cr
f_1&={I_2 \over 2 I_1^2} , \cr
f_2&=-{5 I_2^3 \over 8 I_1^5}+{2 I_2 I_3 \over 3 I_1^4}-
      {I_4 \over 8 I_1^3}, \cr
}
\eqn\polorder
$$
where the first equation is the $g=0$ polymer equation.
With the aid of \momrelation\ these expressions are integrated, yielding
$$
\eqalign{
F_0&=\sum_{n=0}^{\infty} t_{n} {f_0^{n+1} \over (n+1)!},\cr
F_1&=-{1 \over 2} \log I_1, \cr
F_2&= {{-5\,{I_2}^2}\over {24\,{I_1}^3}} +
         {{I_3}\over {8\,{I_1}^2}}. \cr
}
\eqn\polfree
$$

The results suggest that $1/I_1$ can be interpreted as
a propagator and $I_k$ as a $(k+1)$-point
vertex. Thus our ``genus'' expansion is nothing but the loop expansion
and $\lambda$ plays a role of the Planck constant. In fact the explicit
form of $Z$ is obtained in [\NY] which admits the loop expansion in
agreement with ours. We shall now turn to this issue.

It is seen from \polrecursion\ that the polymer
equation is written as [\NY]
$$
t_0+\sum_{n=1}^\infty {t_n \over n!}
\Big(\lambda {\partial \over \partial t_0}+f \Big)^n \cdot 1=0.
\eqno\eq
$$
Substituting $f=\lambda (\log Z)'$ this turns out to be the linear equation
$$
\Big[t_0+\sum_{n=1}^\infty {t_n \over n!}
\Big(\lambda {\partial \over \partial t_0} \Big)^{n} \Big] Z=0,
\eqn\lineareq
$$
which can be easily integrated. We obtain [\NY]
$$
Z=\int dz \exp \Big( {1 \over \lambda}
\sum_{n=0}^{\infty} t_n {z^{n+1} \over (n+1)!} \Big)
 =\int dz \exp {1 \over \lambda} S_{\rm eff}(z).
\eqno\eq
$$
\lineareq\ is then recognized as the Schwinger-Dyson equation for
this integral. It is amusing to note here the similarity of the polymer
action $S_{\rm eff}(z)$  to the gravity action $S_{A_1}$ in \Atwoaction.
Furthermore
the saddle point equation for $Z$ becomes the $g=0$ polymer
equation in \polorder.
Then the perturbation expansion around the classical solution,
$\xi=z-z_{\rm cl}$ with $z_{\rm cl}=f_0$, yields
$$
Z=e^{{1 \over \lambda} S_{\rm eff}(z_{\rm cl})} \int d\xi \exp
  \Big[{1 \over \lambda}
   \Big({1 \over 2} I_1 \xi^2+{1 \over 6} I_2 \xi^3
+{1 \over 24} I_3 \xi^4+\cdots \Big) \Big].
\eqno\eq
$$
Note that $S_{\rm eff}(z_{\rm cl})=F_0$ in \polfree .
Using the Gaussian integration formula
$$
\int d\xi \exp \Big( {1 \over 2 \lambda} I_1 \xi^2 \Big) \xi^{2n}=
\Big( {-2 \pi \lambda \over I_1} \Big)^{1/2} (2n-1)!!
\Big( {-\lambda \over I_1} \Big)^n,
\eqno\eq
$$
one obtains
 $$
\eqalign{
Z=&\Big( {-2 \pi \lambda \over I_1} \Big)^{1/2} e^{{1 \over \lambda}
S_{\rm eff}(z_{\rm cl})}
  \Big[ 1 + \lambda \,\Big( {{-5\,{I_2}^2}\over {24\,{I_1}^3}} +
             {{I_3}\over {8\,{I_1}^2}} \Big)
\cr &
   +{\lambda^{2}}\,\Big( {{385\,{I_2}^4}\over {1152\,{I_1}^6}} -
            {{35\,{I_2}^2\,I_3}\over {64\,{I_1}^5}} +
            {{35\,{I_3}^2}\over {384\,{I_1}^4}} +
            {{7\,I_2\,I_4}\over {48\,{I_1}^4}} -
            {{I_5}\over {48\,{I_1}^3}}
       \Big)
\cr &
   +{\lambda^{3}}\, \Big( {{-85085\,{I_2}^6}\over {82944\,{I_1}^9}} +
            {{25025\,{I_2}^4\,I_3}\over {9216\,{I_1}^8}} -
            {{5005\,{I_2}^2\,{I_3}^2}\over {3072\,{I_1}^7}} +
            {{385\,{I_3}^3}\over {3072\,{I_1}^6}}
\cr &
{}~~~~~~~     -{{1001\,{I_2}^3\,I_4}\over {1152\,{I_1}^7}} +
      {{77\,I_2\,I_3\,I_4}\over {128\,{I_1}^6}} -
      {{21\,{I_4}^2}\over {640\,{I_1}^5}} +
      {{77\,{I_2}^2\,I_5}\over {384\,{I_1}^6}}
\cr &
{}~~~~~~~     -{{7\,I_3\,I_5}\over {128\,{I_1}^5}} -
      {{I_2\,I_6}\over {32\,{I_1}^5}} +
      {{I_7}\over {384\,{I_1}^4}}
       \Big) + \cdots \Big].
}
\eqno\eq
$$
Finally, up to an additive constant $\big( 1/2 \log (-2 \pi \lambda) \big)$,
the free energy is given by
$$
\eqalign{
F=&{1 \over \lambda} S_{\rm eff}(z_{\rm cl})-{1 \over 2} \log I_1
   +\lambda \, \Big( {{-5\,{I_2}^2}\over {24\,{I_1}^3}} +
         {{I_3}\over {8\,{I_1}^2}} \Big)  \cr
 & +  {\lambda^{2}}\, \Big( {{5\,{I_2}^4}\over {16\,{I_1}^6}} -
            {{25\,{I_2}^2\,I_3}\over {48\,{I_1}^5}} +
            {{{I_3}^2}\over {12\,{I_1}^4}} +
            {{7\,I_2\,I_4}\over {48\,{I_1}^4}} -
            {{I_5}\over {48\,{I_1}^3}} \Big) \cr
& +   {\lambda^{3}}\, \Big( {{-1105\,{I_2}^6}\over {1152\,{I_1}^9}} +
            {{985\,{I_2}^4\,I_3}\over {384\,{I_1}^8}} -
            {{445\,{I_2}^2\,{I_3}^2}\over {288\,{I_1}^7}} +
            {{11\,{I_3}^3}\over {96\,{I_1}^6}} -
            {{161\,{I_2}^3\,I_4}\over {192\,{I_1}^7}} \cr
&~~~~~~ +{{7\,I_2\,I_3\,I_4}\over {12\,{I_1}^6}} -
      {{21\,{I_4}^2}\over {640\,{I_1}^5}} +
      {{113\,{I_2}^2\,I_5}\over {576\,{I_1}^6}} -
      {{5\,I_3\,I_5}\over {96\,{I_1}^5}} -
      {{I_2\,I_6}\over {32\,{I_1}^5}} +
      {{I_7}\over {384\,{I_1}^4}} \Big)+\cdots . \cr}
\eqno\eq
$$
The result indeed agrees with \polfree .

Now, a natural question arises: is it possible to find an analogous
integral representation for the partition function in  topological
string theories? If so, what is the physical meaning of such a representation?
In branched polymers $S_{\rm eff}(z)$ is the Landau-Ginzburg action
containing precisely the effect of the fluctuations around the mean field.
Thus what could be imagined in the case of gravity will
be the Landau-Ginzburg type
effective theory for gravity. For pure gravity Yoneya has proposed an
interesting formula for the partition function [\Y]. To write down this
formula explicitly one has to solve the highest weight condition of the
Virasoro algebra, which unfortunately is a difficult task. At present
we do not know how to deal with this problem, though we expect that
deeper understanding of this issue may shed a new light toward the
construction of topological string field theory.

\vskip15mm \fourteenpoint
\centerline{Acknowledgments}

\twelvepoint
The work of T.E. and S.-K.Y. is supported in part by Grant-in-Aid for
Scientific Research on Priority Area 231 ``Infinite Analysis'', Japan
Ministry of Education.
Y.Y. would like to thank T. Kawai and N. Ishibashi for useful
discussions.

\endpage

\Appendix{A}

\undertext{$A_1$ potentials (KdV equations)}

Some constitutive relations for $A_1$ are
$$
\eqalign{
\langle P P \rangle=&R_1=u,\cr
\langle P \sigma_1 \rangle=&R_2={{{u^2}}\over 2} + {{\lambda^2 u''}\over
{12}},\cr
\langle P \sigma_2 \rangle=&R_3={{{u^3}}\over 6} + \lambda^2 [ {{{{u'}^2}}\over
{24}} +
        {{u u''}\over {12}} ]  + {{{\lambda^4} u^{(4)}}\over {240}},\cr
\langle P \sigma_3 \rangle=&R_4
       ={{{u^4}}\over {24}} + \lambda^2 [ {{u {{u'}^2}}\over {24}} +
        {{{u^2} u''}\over {24}} ]  +
     {\lambda^4} [ {{{{u''}^2}}\over {160}} + {{u' u^{(3)}}\over {120}} +
        {{u u^{(4)}}\over {240}} ]  + {{{\lambda^6} u^{(6)}}\over {6720}},\cr
\langle P \sigma_4 \rangle=&R_5={{{u^5}}\over {120}} + \lambda^2
      [ {{{u^2} {{u'}^2}}\over {48}} + {{{u^3} u''}\over {72}} ]
\cr &
       + {\lambda^4} [ {{11 {{u'}^2} u''}\over {1440}} +
       {{u {{u''}^2}}\over {160}} + {{u u' u^{(3)}}\over {120}} +
       {{{u^2} u^{(4)}}\over {480}} ]
\cr &
     + {\lambda^6} [ {{23 {{u^{(3)}}^2}}\over {40320}} +
        {{19 u'' u^{(4)}}\over {20160}} + {{u' u^{(5)}}\over {2240}} +
        {{u u^{(6)}}\over {6720}} ]  + {{{\lambda^8} u^{(8)}}\over {241920}}
    ,\cr
\langle \sigma_1 \sigma_1 \rangle=&
        {{{u^3}}\over 3} + \lambda^2 [ {{{{u'}^2}}\over {24}} +
        {{u u''}\over 6} ]  + {{{\lambda^4} u^{(4)}}\over {144}},\cr
\langle \sigma_1 \sigma_2 \rangle=&
        {{{u^4}}\over 8} + \lambda^2 [ {{u {{u'}^2}}\over {12}} +
        {{{u^2} u''}\over 8} ]
\cr &
     + {\lambda^4} [ {{23 {{u''}^2}}\over {1440}} +
        {{u' u^{(3)}}\over {60}} + {{u u^{(4)}}\over {90}} ]  +
     {{{\lambda^6} u^{(6)}}\over {2880}},\cr
\langle \sigma_2 \sigma_2 \rangle=&{{{u^5}}\over {20}} + \lambda^2
      [ {{{u^2} {{u'}^2}}\over {12}} + {{{u^3} u''}\over {12}} ]
\cr &
       + {\lambda^4} [ {{7 {{u'}^2} u''}\over {240}} +
        {{23 u {{u''}^2}}\over {720}} + {{u u' u^{(3)}}\over {30}} +
        {{{u^2} u^{(4)}}\over {90}} ]
\cr &
     + {\lambda^6} [ {{13 {{u^{(3)}}^2}}\over {5760}} +
        {{u'' u^{(4)}}\over {240}} + {{u' u^{(5)}}\over {576}} +
        {{u u^{(6)}}\over {1440}} ]  + {{{\lambda^8} u^{(8)}}\over {57600}}
    .\cr
}
\eqno\eq$$

In general, $R_n$ is determined by the Gelfand-Dikii recursion relation [\GD]\
$$
(2n+1) D R_{n+1}=[{\lambda^2 \over 4} D^3+2u D+u'] R_n,
\eqno\eq
$$
and we have the genus expansion of $R_n$
$$
R_n=\sum_{g=0}^{\infty} \lambda^{2g} R_{n ,g}, \hskip 1cm
R_{n ,g}=\sum_{k=g+1}^{3g} {u^{n-k} \over (n-k)!}
P_{n ,g}(u',\cdots, u^{(2g)}),
\eqno\eq$$
where $P_{n ,g}$ is a polynomial in $u',u'',\cdots, u^{(2g)}$.

Explicit formulas up to $g=5$ are given as follows:
$$
R_{n,0}={u^n \over n!},
\eqno\eq$$
$$
R_{n,1}={1 \over 12} {u^{n-2} \over (n-2)!} u''
+{1 \over 24} {u^{n-3} \over (n-3)!}  u'^2,
\eqno\eq$$
$$\eqalign{
R_{n,2}&=
   {u^{n-6} \over (n-6)!} {{u'}^4\over 1152} +
   {u^{n-5} \over (n-5)!} {11 {u'}^2 u'' \over 1440}
\cr &
   + {u^{n-4} \over (n-4)!} [ {{{{u''}^2}}\over {160}}
   + {{u' u^{(3)}}\over {120}} ]
   + {u^{n-3} \over (n-3)!} {u^{(4)}\over 240}
,}
\eqno\eq$$
$$\eqalign{
R_{n,3}&=
   {u^{n-9} \over (n-9)!} {{u'}^6\over 82944} +
   {u^{n-8} \over (n-8)!} {17 {u'}^4 u'' \over 69120} +
   {u^{n-7} \over (n-7)!} [ {{83 {{u'}^2} {{u''}^2}}\over {80640}} +
      {{{{u'}^3} u^{(3)}}\over {2016}} ]
\cr &
   +{u^{n-6} \over (n-6)!} [ {{61 {{u''}^3}}\over {120960}} +
    {{43 u' u'' u^{(3)}}\over {20160}} +
    {{5 {{u'}^2} u^{(4)}}\over {8064}} ]  +
\cr &
   +{u^{n-5} \over (n-5)!} [ {{23 {{u^{(3)}}^2}}\over {40320}}
    {{19 u'' u^{(4)}}\over {20160}} + {{u' u^{(5)}}\over {2240}} ]  +
    {u^{n-4} \over (n-4)!} {u^{(6)}\over 6720}
,}
\eqno\eq$$
$$\eqalign{
R_{n,4}&=
   {u^{n-12} \over (n-12)!} {{u'}^8 \over 7962624} +
   {u^{n-11} \over (n-11)!} {23 {u'}^6 u''\over 4976640} +
   {u^{n-10} \over (n-10)!} [ {{893 {{u'}^4} {{u''}^2}}\over {19353600}} +
      {{13 {{u'}^5} u^{(3)}}\over {967680}} ]
\cr &
   +  {u^{n-9} \over (n-9)!} [ {{259 {{u'}^2} {{u''}^3}}\over {2073600}} +
      {{439 {{u'}^3} u'' u^{(3)}}\over {2419200}} +
      {{17 {{u'}^4} u^{(4)}}\over {645120}} ]
\cr &
   +  {u^{n-8} \over (n-8)!} [ {{1261 {{u''}^4}}\over {29030400}} +
      {{227 u' {{u''}^2} u^{(3)}}\over {604800}} +
      {{659 {{u'}^2} {{u^{(3)}}^2}}\over {4838400}} +
      {{527 {{u'}^2} u'' u^{(4)}}\over {2419200}} +
      {{17 {{u'}^3} u^{(5)}}\over {483840}} ]
\cr &
   +  {u^{n-7} \over (n-7)!} [ {{31 u'' {{u^{(3)}}^2}}\over {161280}} +
      {{5 {{u''}^2} u^{(4)}}\over {32256}} +
      {{u' u^{(3)} u^{(4)}}\over {4480}} + {{u' u'' u^{(5)}}\over {6720}} +
      {{{{u'}^2} u^{(6)}}\over {32256}} ]
\cr &
   +  {u^{n-6} \over (n-6)!} [ {{23 {{u^{(4)}}^2}}\over {483840}} +
      {{19 u^{(3)} u^{(5)}}\over {241920}} + {{11 u'' u^{(6)}}\over {241920}} +
      {{u' u^{(7)}}\over {60480}} ]
\cr &
   + {u^{n-5} \over (n-5)!} {u^{(8)}\over 241920}
,}
\eqno\eq$$
\vfill \break
$$\eqalign{
R_{n,5}&=
   {u^{n-15} \over (n-15)!} {{u'}^{10}\over 955514880} +
   {u^{n-14} \over (n-14)!} {29 {u'}^8 u''\over 477757440} +
   {u^{n-13} \over (n-13)!} [ {{19 {{u'}^6} {{u''}^2}}\over {17203200}} +
      {{{{u'}^7} u^{(3)}}\over {4354560}} ]
\cr &
   +  {u^{n-12} \over (n-12)!} [ {{187 {{u'}^4} {{u''}^3}}\over {25804800}} +
      {{7 {{u'}^5} u'' u^{(3)}}\over {1105920}} +
      {{17 {{u'}^6} u^{(4)}}\over {27869184}} ]
\cr &
   +  {u^{n-11} \over (n-11)!} [ {{4097 {{u'}^2} {{u''}^4}}\over {283852800}} +
      {{4471 {{u'}^3} {{u''}^2} u^{(3)}}\over {106444800}} +
      {{1303 {{u'}^4} {{u^{(3)}}^2}}\over {170311680}} +
      {{23 {{u'}^4} u'' u^{(4)}}\over {1892352}} +
      {{299 {{u'}^5} u^{(5)}}\over {255467520}} ]
\cr &
   +  {u^{n-10} \over (n-10)!} [ {{79 {{u''}^5}}\over {20275200}} +
      {{2579 u' {{u''}^3} u^{(3)}}\over {45619200}} +
      {{1877 {{u'}^2} u'' {{u^{(3)}}^2}}\over {30412800}} +
      {{1493 {{u'}^2} {{u''}^2} u^{(4)}}\over {30412800}}
\cr & \hskip 5mm
    + {{127 {{u'}^3} u^{(3)} u^{(4)}}\over {5322240}} +
      {{839 {{u'}^3} u'' u^{(5)}}\over {53222400}} +
      {{139 {{u'}^4} u^{(6)}}\over {85155840}} ]
\cr &
   +  {u^{n-9} \over (n-9)!} [ {{3851 {{u''}^2} {{u^{(3)}}^2}}\over {91238400}}
+
      {{467 u' {{u^{(3)}}^3}}\over {22809600}} +
      {{7171 {{u''}^3} u^{(4)}}\over {319334400}} +
      {{15629 u' u'' u^{(3)} u^{(4)}}\over {159667200}}
\cr & \hskip 5mm
    + {{1817 {{u'}^2} {{u^{(4)}}^2}}\over {127733760}} +
      {{383 u' {{u''}^2} u^{(5)}}\over {11827200}} +
      {{7543 {{u'}^2} u^{(3)} u^{(5)}}\over {319334400}} +
      {{4283 {{u'}^2} u'' u^{(6)}}\over {319334400}} +
      {{13 {{u'}^3} u^{(7)}}\over {7983360}} ]
\cr &
   +  {u^{n-8} \over (n-8)!} [ {{7939 {{u^{(3)}}^2} u^{(4)}}\over {319334400}}
+
      {{6353 u'' {{u^{(4)}}^2}}\over {319334400}} +
      {{13 u'' u^{(3)} u^{(5)}}\over {394240}} +
      {{3067 u' u^{(4)} u^{(5)}}\over {159667200}}
\cr & \hskip 5mm
   +  {{3001 {{u''}^2} u^{(6)}}\over {319334400}} +
      {{13 u' u^{(3)} u^{(6)}}\over {950400}} +
      {{109 u' u'' u^{(7)}}\over {15966720}} +
      {{71 {{u'}^2} u^{(8)}}\over {63866880}} ]
\cr &
   +  {u^{n-7} \over (n-7)!} [ {{71 {{u^{(5)}}^2}}\over {21288960}} +
      {{61 u^{(4)} u^{(6)}}\over {10644480}} +
      {{19 u^{(3)} u^{(7)}}\over {5322240}} +
      {{17 u'' u^{(8)}}\over {10644480}} + {{u' u^{(9)}}\over {2128896}}
       ]
\cr &
     + {u^{n-6} \over (n-6)!} {u^{(10)}\over 10644480}.
}
\eqno\eq$$

\endpage

\Appendix{B}

\undertext{$A_2$ potentials (Boussinesq equation)}

The Boussinesq potentials $(R_n, S_n)$ are obtained by the following
recursion relation:
$$\eqalign{
(n+3) D R_{n+3}&=
(3v D+2 v')R_n+({2 \lambda^2 \over 3} D^3 +2 u D+u')S_n, \cr
(n+3) D S_{n+3}&=
[{\lambda^4  \over 18} D^5+{5 \lambda^2 \over 6} u D^3+{5 \lambda^2 \over 4} u'
D^2
    +({3 \lambda^2 \over 4} u''+2 u^2) D+({\lambda^2 \over 6} u'''+2 u u')]R_n
\cr
       &+ (3 v D+v')S_n, \cr
}
\eqno\eq
$$
where $'=\partial /\partial t_{0,0}=D$.

Initial data are
$$
\eqalign{
\langle PP \rangle=R_1=u, \hskip 1cm
&\langle PQ \rangle=S_1=v, \hskip 1cm
\langle QP \rangle=R_2=v, \cr
 \hskip 1cm &\langle QQ \rangle=S_2={u^2 \over 2}+\lambda^2 {u'' \over 12}.
}
\eqno\eq
$$
Hence we have
$$
\langle \sigma_1(P)P \rangle=
R_4= u v + \lambda^2 {{v''}\over 6},
\eqno\eq
$$
$$
\langle \sigma_1(P)Q \rangle=
S_4= {{{{u}^3}}\over 3} + {{{{v}^2}}\over 2}
 + \lambda^2 [{{{{u'}^2}}\over 8} + {{u u''}\over 4} ]
 + \lambda^4  {{u^{(4)}}\over {72}},
\eqno\eq
$$
$$
\langle \sigma_1(Q)P \rangle=
R_5= {{{{u}^3}}\over 6} + {{{{v}^2}}\over 2}
 + \lambda^2 [ {{{{u'}^2}}\over 8} + {{u u''}\over 6} ]
 + \lambda^4  {{u^{(4)}}\over {90}},
\eqno\eq
$$
$$
\langle \sigma_1(Q)Q \rangle=
S_5= {{{{u}^2} v }\over 2}
 + \lambda^2 [{{u' v'}\over {12}} +{{v u''}\over {12}} + {{u v''}\over 6} ]
 + \lambda^4  {{v^{(4)}}\over {90}},
\eqno\eq$$
$$
\eqalign{
\langle \sigma_2(P)P \rangle=
R_7&= {{{{u}^4}}\over {12}} + {{u {{v}^2}}\over 2}
 + \lambda^2 [{{5 u {{u'}^2}}\over {24}} + {{{{v'}^2}}\over {12}} +
   {{{{u}^2} u''}\over 6} +{{v v''}\over 6}]
\cr &
 + \lambda^4  [{{7 {{u''}^2}}\over {144}} +
   {{5 u' u^{(3)}}\over {72}} +
   {{u u^{(4)}}\over {36}} ]
 + \lambda^6 {{u^{(6)}}\over {756}},
}
\eqno\eq$$
$$
\eqalign{
\langle \sigma_2(P)Q \rangle=
S_7&={{{{u}^3} v }\over 3} + {{{{v}^3}}\over 6}
 + \lambda^2 [ {{v {{u'}^2}}\over 8} + {{u u' v'}\over 4} +
   {{u v u''}\over 4} + {{{{u}^2} v''}\over 6} ]
\cr &
 + \lambda^4  [{{u'' v''}\over {18}} + {{v' u^{(3)}}\over {36}} +
   {{u' v^{(3)}}\over {24}} + {{v u^{(4)}}\over {72}} +
   {{u v^{(4)}}\over {36}}] +  \lambda^6 {{v^{(6)}}\over {756}},
}
\eqno\eq$$
$$
\eqalign{
\langle \sigma_2(Q)P \rangle=
R_8&= {{{{u}^3} v }\over 6} + {{{{v}^3}}\over 6}
 + \lambda^2 [{{v {{u'}^2}}\over 8} + {{u u' v'}\over 6} +
   {{u v u''}\over 6} + {{{{u}^2} v''}\over {12}} ]
\cr &
 + \lambda^4  [{{13 u'' v''}\over {360}} + {{7 v' u^{(3)}}\over {360}} +
   {{u' v^{(3)}}\over {30}} + {{v u^{(4)}}\over {90}} +
   {{u v^{(4)}}\over {60}} ]+ \lambda^6 {{v^{(6)}}\over {1080}},
}
\eqno\eq$$

and
$$
\eqalign{
\langle \sigma_2(Q)Q \rangle & =
S_8= {{{{u}^5}}\over {30}} + {{{{u}^2} {{v}^2}}\over 4}
   + \lambda^2 [{{7 {{u}^2} {{u'}^2}}\over {48}} +
   {{v u' v'}\over {12}} + {{u {{v'}^2}}\over {12}} +
   {{7 {{u}^3} u''}\over {72}} + {{{{v}^2} u''}\over {24}}
   + {{u v v''}\over 6} ]
\cr &
  + \lambda^4  [ {{11 {{u'}^2} u''}\over {160}} +
   {{17 u {{u''}^2}}\over {240}} +
   {{{{v''}^2}}\over {45}} + {{17 u u' u^{(3)}}\over {180}} +
   {{v' v^{(3)}}\over {40}} +
   {{17 {{u}^2} u^{(4)}}\over {720}}+ {{v v^{(4)}}\over {90}} ]
\cr &
  + \lambda^6 [{{7 {{u^{(3)}}^2}}\over {720}} +{{67 u'' u^{(4)}}\over {4320}} +
   {{u' u^{(5)}}\over {144}} + {{u u^{(6)}}\over {432}}] +
   \lambda^8 {{u^{(8)}}\over {12960}}.
}
\eqno\eq$$

Next, some two-point functions of descendants read as follows.
$$
\eqalign{
  \langle \sigma_1(P) \sigma_1(P) \rangle &=
  {{{{u}^4}}\over 4} + u {{v}^2}
  + \lambda^2 [{{u {{u'}^2}}\over 3}
   +{{{{v'}^2}}\over {12}} + {{5 {{u}^2} u''}\over {12}} +
   {{13 {{u''}^2}}\over {144}} + {{v v''}\over 3} ]
\cr &
  + \lambda^4  [{{13 {{u''}^2}}\over {144}} +{{u' u^{(3)}}\over 9}
    + {{u u^{(4)}}\over {18}} ]+
    \lambda^6 {{u^{(6)}}\over {432}},
}
\eqno\eq$$
$$
\eqalign{
 \langle \sigma_1(Q) \sigma_1(P) \rangle &=
   {{{{u}^3} v }\over 2} + {{{{v}^3}}\over 3}
  + \lambda^2 [{{v {{u'}^2}}\over 4} + {{u u' v'}\over 4} +
   {{5 u v u''}\over {12}} + {{{{u}^2} v''}\over 4} ]
\cr &
  + \lambda^4  [{{29 u'' v''}\over {360}} + {{11 v' u^{(3)}}\over {360}} +
   {{7 u' v^{(3)}}\over {120}} + {{v u^{(4)}}\over {40}} +
   {{7 u v^{(4)}}\over {180}}] + \lambda^6 {{v^{(6)}}\over {540}},
}
\eqno\eq$$
$$
\eqalign{
  \langle \sigma_2(P) \sigma_1(P) \rangle &=
   {{{{u}^4} v }\over 3} +
   {{u {{v}^3}}\over 2}
\cr &
+ \lambda^2 [{{13 u v {{u'}^2}}\over {24}} +
   {{5 {{u}^2} u' v'}\over {12}} + {{v {{v'}^2}}\over 6} +
   {{7 {{u}^2} v u''}\over {12}}+ {{2 {{u}^3} v''}\over 9} +
   {{{{v}^2} v''}\over 4}]
\cr &
+ \lambda^4  [{{17 u' v' u''}\over {72}} +
   {{5 v {{u''}^2}}\over {36}} + {{7 {{u'}^2} v''}\over {48}} +
   {{19 u u'' v''}\over {72}} +
   {{13 v u' u^{(3)}}\over {72}}
\cr & \hskip 5mm
   + {{u v' u^{(3)}}\over 8} + {{u u' v^{(3)}}\over 6}
   +{{u v u^{(4)}}\over {12}} +{{{{u}^2} v^{(4)}}\over {18}}]
\cr &
+ \lambda^6 [ {{29 u^{(3)} v^{(3)}}\over {1008}} +
   {{67 v'' u^{(4)}}\over {3024}} +
   {{11 u'' v^{(4)}}\over {378}} + {{v' u^{(5)}}\over {126}} +
   {{5 u' v^{(5)}}\over {336}}
\cr & \hskip 5mm
   + {{11 v u^{(6)}}\over {3024}} + {{u v^{(6)}}\over {168}}]
   + \lambda^8 {{v^{(8)}}\over {4536}}.
}
\eqno\eq$$
\endpage

\Appendix{C}

\undertext{$D_4$ potentials}

There are four primaries $P,Q,R$ and $S$ in the $D_4$ model.
The order parameters are defined by
$$
   u=\langle { P} { P} \rangle,
 \ v=\langle { P} { Q} \rangle,
 \ w=\langle { P} { R} \rangle,
 \ f=\langle { P} { S} \rangle.
\eqno\eq$$

We determine the constitutive relations by requiring the commutativity of
flow equations. Starting with the dispersionless ($\lambda =0$) terms we find
$$
\eqalign{
\langle P Q \rangle =& v, \cr
\langle Q Q \rangle =&{{{u^3}}\over 3} + u v + w +
   \lambda^2 [ {{u u''}\over 6} + {{v''}\over 6} ]  +
   {{{\lambda^4 } u^{(4)}}\over {180}}, \cr
\langle R Q \rangle =&{{{f^2}}\over 2} + {u^2} v + {{{v^2}}\over 2} +
   \lambda^2 [ {{u' v'}\over 2} + {{v u''}\over 3} + {{u v''}\over 2} ]
     + {{{\lambda^4 } v^{(4)}}\over {20}}, \cr
\langle S Q \rangle =& -f u  - {{\lambda^2 f''}\over 6}, \cr
\langle P R \rangle =&w
,\cr
\langle Q R \rangle =&{{{f^2}}\over 2} + {u^2} v + {{{v^2}}\over 2} +
   \lambda^2 [ {{u' v'}\over 2} + {{v u''}\over 3} + {{u v''}\over 2} ]
     + {{{\lambda^4 } v^{(4)}}\over {20}}
,\cr
\langle R R \rangle =&- {f^2} u   + {{{u^5}}\over 5} + u {v^2}
\cr &
    + \lambda^2 [ -{{{{f'}^2}}\over {12}} + {u^2} {{u'}^2} +
      {{{{v'}^2}}\over {12}} + {{u' w'}\over 6} - {{f f''}\over 3} +
      {{2 {u^3} u''}\over 3} + {{v v''}\over 3} + {{u w''}\over 3} ]
\cr &
     + {\lambda^4 } [ {{11 {{u'}^2} u''}\over {18}} +
      {{5 u {{u''}^2}}\over {12}} + {{5 u u' u^{(3)}}\over 9} +
      {{5 {u^2} u^{(4)}}\over {36}} + {{2 w^{(4)}}\over {45}} ]
\cr &
     + {\lambda^6 } [ {{7 {{u^{(3)}}^2}}\over {216}} +
      {{13 u'' u^{(4)}}\over {216}} + {{u' u^{(5)}}\over {30}} +
      {{u u^{(6)}}\over {90}} ]  + {{{\lambda^8} u^{(8)}}\over {3600}}
,\cr
\langle S R \rangle =&f {u^2} - f v + \lambda^2 [ {{f' u'}\over 2} +
      {{u f''}\over 2} +
      {{f u''}\over 3} ]  + {{{\lambda^4 } f^{(4)}}\over {20}}
,\cr
\langle P S \rangle =&f
,\cr
\langle Q S \rangle =& -f u  - {{\lambda^2 f''}\over 6}
,\cr
\langle R S \rangle =& f {u^2} - f v + \lambda^2 [ {{f' u'}\over 2} +
      {{u f''}\over 2} +
      {{f u''}\over 3} ]  + {{{\lambda^4 } f^{(4)}}\over {20}}
,\cr
\langle S S \rangle =& -{{{u^3}}\over 3} + u v - w +
   \lambda^2 [ -{{u u'' }\over 6} + {{v''}\over 6} ]  -
   {{{\lambda^4 } u^{(4)}}\over {180}}
,\cr
}
\eqno\eq$$
\vfill \break
$$
\eqalign{
\langle P \sigma_1 (P) \rangle =&-{{{f^2}}\over 2} + {{{v^2}}\over 2} + u w +
   \lambda^2 [ {{u {{u'}^2}}\over 2} + {{{u^2} u''}\over 6} +
      {{w''}\over 3} ]
\cr &
   + {\lambda^4 } [ {{{{u''}^2}}\over {36}} + {{u' u^{(3)}}\over {12}} +
      {{u u^{(4)}}\over {30}} ]  + {{{\lambda^6 } u^{(6)}}\over {840}}
,\cr
\langle Q \sigma_1 (P) \rangle =&{f^2} u + {u^3} v + u {v^2} + v w
\cr &
   + \lambda^2 [ {{{{f'}^2}}\over 4} + {{v {{u'}^2}}\over 2} + 2 u u' v' +
      {{{{v'}^2}}\over 4} + {{f f''}\over 2} + {{7 u v u''}\over 6} +
      {u^2} v'' + {{v v''}\over 2} ]
\cr &
   + {\lambda^4 } [ {{2 u'' v''}\over 3} + {{5 v' u^{(3)}}\over {12}} +
      {{u' v^{(3)}}\over 2} + {{7 v u^{(4)}}\over {60}} +
      {{u v^{(4)}}\over 4} ]  + {{{\lambda^6 } v^{(6)}}\over {56}}
,\cr
\langle R \sigma_1 (P) \rangle =&-{{3 {f^2} {u^2}}\over 2} + {{{u^6}}\over 6} +
    {f^2} v +
   {{3 {u^2} {v^2}}\over 2} + {{{v^3}}\over 3} + {{{w^2}}\over 2}
\cr &
   + \lambda^2 [ -{{u {{f'}^2}}\over 2} -
      {{3 f f' u'}\over 2} + 2 {u^3} {{u'}^2} +
      {{3 v u' v'}\over 2} + {{u {{v'}^2}}\over 2} +
      {{u u' w'}\over 2} - {{3 f u f''}\over 2}
\cr &
      -{{2 {f^2} u''}\over 3} + {u^4} u'' + {{2 {v^2} u''}\over 3} +
      {{3 u v v''}\over 2} + {{{u^2} w''}\over 2} ]
\cr &
   +{\lambda^4 } [ {{7 {{u'}^4}}\over {12}} - {{23 {{f''}^2}}\over {120}} +
      5 u {{u'}^2} u'' + {{131 {u^2} {{u''}^2}}\over {72}} +
      {{23 {{v''}^2}}\over {120}} + {{49 u'' w''}\over {180}}
\cr &
      -{{41 f' f^{(3)}}\over {180}} + {{29 {u^2} u' u^{(3)}}\over {12}} +
      {{4 w' u^{(3)}}\over {45}} + {{41 v' v^{(3)}}\over {180}} +
      {{17 u' w^{(3)}}\over {60}} - {{29 f f^{(4)}}\over {180}}
\cr &
      +{{13 {u^3} u^{(4)}}\over {36}} + {{29 v v^{(4)}}\over {180}} +
      {{17 u w^{(4)}}\over {90}} ]
\cr &
    + {\lambda^6 } [ {{1799 {{u''}^3}}\over {3240}} +
      {{2491 u' u'' u^{(3)}}\over {1080}} +
      {{529 u {{u^{(3)}}^2}}\over {1080}}
\cr &
      +{{29 {{u'}^2} u^{(4)}}\over {45}} +
      {{877 u u'' u^{(4)}}\over {1080}} +
      {{53 u u' u^{(5)}}\over {135}} + {{31 {u^2} u^{(6)}}\over {540}} +
      {{121 w^{(6)}}\over {7560}} ]
\cr &
    + {\lambda^8} [ {{301 {{u^{(4)}}^2}}\over {7200}} +
      {{2351 u^{(3)} u^{(5)}}\over {32400}} +
      {{1499 u'' u^{(6)}}\over {32400}}
\cr &
      +{{197 u' u^{(7)}}\over {10800}} + {{43 u u^{(8)}}\over {10800}}
       ]  + {{{\lambda^{10}} u^{(10)}}\over {10800}}
,\cr
\langle S \sigma_1 (P) \rangle =&f {u^3} - 2 f u v + f w
\cr &
   + \lambda^2 [ 2 u f' u' + {{f {{u'}^2}}\over 2} - {{f' v'}\over 2} +
      {u^2} f'' - {{v f''}\over 2} + {{7 f u u''}\over 6} -
      {{f v''}\over 2} ]
\cr &
    + {\lambda^4 } [ {{2 f'' u''}\over 3} + {{u' f^{(3)}}\over 2} +
      {{5 f' u^{(3)}}\over {12}} + {{u f^{(4)}}\over 4} +
      {{7 f u^{(4)}}\over {60}} ]  + {{{\lambda^6 } f^{(6)}}\over {56}}
.\cr
}
\eqno\eq$$
\endpage

\Appendix{D}

\undertext{$A_1$ free energy (1)}

$$
\eqalign{
F_1&={1 \over {24}} \log u',
\cr
F_2&=
   {{{{u''}^3}}\over {360 {{u'}^4}}} -
   {{7 u'' u^{(3)}}\over {1920 {{u'}^3}}} +
   {{u^{(4)}}\over {1152 {{u'}^2}}},
\cr
F_3&=
   -{{5 {{u''}^6}}\over {648 {{u'}^8}}} +
   {{59 {{u''}^4} u^{(3)}}\over {3024 {{u'}^7}}} -
   {{83 {{u''}^2} {{u^{(3)}}^2}}\over {7168 {{u'}^6}}} +
   {{59 {{u^{(3)}}^3}}\over {64512 {{u'}^5}}}
\cr &
   -{{83 {{u''}^3} u^{(4)}}\over {15120 {{u'}^6}}} +
   {{1273 u'' u^{(3)} u^{(4)}}\over {322560 {{u'}^5}}} -
   {{103 {{u^{(4)}}^2}}\over {483840 {{u'}^4}}} +
   {{353 {{u''}^2} u^{(5)}}\over {322560 {{u'}^5}}}
\cr &
   -{{53 u^{(3)} u^{(5)}}\over {161280 {{u'}^4}}} -
   {{7 u'' u^{(6)}}\over {46080 {{u'}^4}}} +
   {{u^{(7)}}\over {82944 {{u'}^3}}},
\cr
F_4&=
   {{463 {{u''}^9}}\over {4860 {{u'}^{12}}}} -
   {{193 {{u''}^7} u^{(3)}}\over {540 {{u'}^{11}}}} +
   {{14903 {{u''}^5} {{u^{(3)}}^2}}\over {34560 {{u'}^{10}}}} -
   {{305129 {{u''}^3} {{u^{(3)}}^3}}\over {1658880 {{u'}^9}}}
\cr &
   +{{22809 u'' {{u^{(3)}}^4}}\over {1146880 {{u'}^8}}} +
   {{619 {{u''}^6} u^{(4)}}\over {6075 {{u'}^{10}}}} -
   {{101503 {{u''}^4} u^{(3)} u^{(4)}}\over {518400 {{u'}^9}}} +
   {{13138507 {{u''}^2} {{u^{(3)}}^2} u^{(4)}}\over {154828800 {{u'}^8}}}
\cr &
   -{{2153 {{u^{(3)}}^3} u^{(4)}}\over {460800 {{u'}^7}}} +
   {{1823 {{u''}^3} {{u^{(4)}}^2}}\over {90720 {{u'}^8}}} -
   {{44201 u'' u^{(3)} {{u^{(4)}}^2}}\over {4423680 {{u'}^7}}} +
   {{229 {{u^{(4)}}^3}}\over {995328 {{u'}^6}}}
\cr &
   -{{2243 {{u''}^5} u^{(5)}}\over {103680 {{u'}^9}}} +
   {{415273 {{u''}^3} u^{(3)} u^{(5)}}\over {13271040 {{u'}^8}}} -
   {{12035 u'' {{u^{(3)}}^2} u^{(5)}}\over {1548288 {{u'}^7}}} -
   {{171343 {{u''}^2} u^{(4)} u^{(5)}}\over {30965760 {{u'}^7}}}
\cr &
   +{{949 u^{(3)} u^{(4)} u^{(5)}}\over {884736 {{u'}^6}}} +
   {{9221 u'' {{u^{(5)}}^2}}\over {30965760 {{u'}^6}}} +
   {{12907 {{u''}^4} u^{(6)}}\over {3628800 {{u'}^8}}} -
   {{60941 {{u''}^2} u^{(3)} u^{(6)}}\over {17203200 {{u'}^7}}}
\cr &
   +{{59 {{u^{(3)}}^2} u^{(6)}}\over {172032 {{u'}^6}}} +
   {{15179 u'' u^{(4)} u^{(6)}}\over {30965760 {{u'}^6}}} -
   {{197 u^{(5)} u^{(6)}}\over {6193152 {{u'}^5}}} -
   {{212267 {{u''}^3} u^{(7)}}\over {464486400 {{u'}^7}}}
\cr &
   +{{20639 u'' u^{(3)} u^{(7)}}\over {77414400 {{u'}^6}}} -
   {{2069 u^{(4)} u^{(7)}}\over {92897280 {{u'}^5}}} +
   {{2323 {{u''}^2} u^{(8)}}\over {51609600 {{u'}^6}}} -
   {{163 u^{(3)} u^{(8)}}\over {15482880 {{u'}^5}}}
\cr &
   -{{7 u'' u^{(9)}}\over {2211840 {{u'}^5}}} +
   {{u^{(10)}}\over {7962624 {{u'}^4}}},
\cr
}
\eqno\eq
$$
where
$$
u^{(n)}={\partial^n \over \partial t_0^n} u,
\hskip1cm
u=\sum_{n=0}^{\infty} t_n {u^n \over n!}.
\eqno\eq
$$

\vfill\break

\undertext{$A_1$ free energy (2)}

$$\eqalign{
F_1&=-{1 \over {24}} \log M
\cr
F_2&={{7 {{G[2]}^3}}\over {1440 {M^5}}} +
   {{29 G[2] G[3]}\over {5760 {M^4}}} + {{G[4]}\over {1152 {M^3}}},
\cr
F_3&={{245 {{G[2]}^6}}\over {20736 {M^{10}}}} +
   {{193 {{G[2]}^4} G[3]}\over {6912 {M^9}}} +
   {{205 {{G[2]}^2} {{G[3]}^2}}\over {13824 {M^8}}} +
   {{583 {{G[3]}^3}}\over {580608 {M^7}}}
\cr &
   +{{53 {{G[2]}^3} G[4]}\over {6912 {M^8}}} +
   {{1121 G[2] G[3] G[4]}\over {241920 {M^7}}} +
   {{607 {{G[4]}^2}}\over {2903040 {M^6}}} +
   {{17 {{G[2]}^2} G[5]}\over {11520 {M^7}}}
\cr &
   +{{503 G[3] G[5]}\over {1451520 {M^6}}} +
   {{77 G[2] G[6]}\over {414720 {M^6}}} + {{G[7]}\over {82944 {M^5}}},
\cr
F_4&={{259553 {{G[2]}^9}}\over {2488320 {M^{15}}}} +
   {{475181 {{G[2]}^7} G[3]}\over {1244160 {M^{14}}}} +
   {{145693 {{G[2]}^5} {{G[3]}^2}}\over {331776 {M^{13}}}} +
   {{43201 {{G[2]}^3} {{G[3]}^3}}\over {248832 {M^{12}}}}
\cr &
   +{{134233 G[2] {{G[3]}^4}}\over {7962624 {M^{11}}}} +
   {{14147 {{G[2]}^6} G[4]}\over {124416 {M^{13}}}} +
   {{83851 {{G[2]}^4} G[3] G[4]}\over {414720 {M^{12}}}} +
   {{26017 {{G[2]}^2} {{G[3]}^2} G[4]}\over {331776 {M^{11}}}}
\cr &
   +{{185251 {{G[3]}^3} G[4]}\over {49766400 {M^{10}}}} +
   {{5609 {{G[2]}^3} {{G[4]}^2}}\over {276480 {M^{11}}}} +
   {{177 G[2] G[3] {{G[4]}^2}}\over {20480 {M^{10}}}} +
   {{175 {{G[4]}^3}}\over {995328 {M^9}}}
\cr &
   +{{21329 {{G[2]}^5} G[5]}\over {829440 {M^{12}}}} +
   {{13783 {{G[2]}^3} G[3] G[5]}\over {414720 {M^{11}}}} +
   {{1837 G[2] {{G[3]}^2} G[5]}\over {259200 {M^{10}}}} +
   {{7597 {{G[2]}^2} G[4] G[5]}\over {1382400 {M^{10}}}}
\cr &
   +{{719 G[3] G[4] G[5]}\over {829440 {M^9}}} +
   {{533 G[2] {{G[5]}^2}}\over {1935360 {M^9}}} +
   {{2471 {{G[2]}^4} G[6]}\over {552960 {M^{11}}}} +
   {{7897 {{G[2]}^2} G[3] G[6]}\over {2073600 {M^{10}}}}
\cr &
   +{{1997 {{G[3]}^2} G[6]}\over {6635520 {M^9}}} +
   {{1081 G[2] G[4] G[6]}\over {2322432 {M^9}}} +
   {{487 G[5] G[6]}\over {18579456 {M^8}}} +
   {{4907 {{G[2]}^3} G[7]}\over {8294400 {M^{10}}}}
\cr &
   +{{16243 G[2] G[3] G[7]}\over {58060800 {M^9}}} +
   {{1781 G[4] G[7]}\over {92897280 {M^8}}} +
   {{53 {{G[2]}^2} G[8]}\over {921600 {M^9}}} +
   {{947 G[3] G[8]}\over {92897280 {M^8}}}
\cr &
   +{{149 G[2] G[9]}\over {39813120 {M^8}}} +
   {{G[10]}\over {7962624 {M^7}}},
\cr}
\eqno\eq
$$
where
$$
\eqalign{
M&=1-G[1],\cr
G[k]&=\sum_{n=0}^{\infty} t_{n+k} {u(t)^n \over n!}.
}
\eqno\eq
$$
The results are in agreement with those obtained previously in
[\IZ,\Amb].


\refout


\bye